\documentclass[journal]{IEEEtran}
\usepackage{epsfig,color}
\usepackage{graphicx}
\usepackage{amsbsy}
\usepackage{amsmath}
\usepackage{amssymb}
\usepackage{euscript}
\usepackage{prodint}
\usepackage{chngcntr}

\newtheorem{myproperty}{\bf Property}
\newtheorem{mytheorem}{\bf Theorem}
\newtheorem{mylemma}{\bf Lemma}

\newtheorem{myremark}{Remark}

\newtheorem{myexample}{\it Example}
\newtheorem{mydefinition}{\bf Definition}

\begin{document}

\title{The Two-Modular Fourier Transform \\ of Binary Functions }
\author{Yi~Hong,~\IEEEmembership{Senior Member,~IEEE,}
        Emanuele~Viterbo,~\IEEEmembership{Fellow,~IEEE,}
        and~Jean-Claude~Belfiore,~\IEEEmembership{Member,~IEEE}
\thanks{Yi Hong and Emanuele Viterbo are with the Department
of Electrical and Computer Systems Engineering, Faculty of Engineering, Monash University, VIC 3800,
e-mail: $\{$yi.hong, emanuele.viterbo$\}$@monash.edu. Jean-Claude~Belfiore is with Communications
and Electronics Dept., Telecom ParisTech, Paris, France, email: jean-claude.belfiore@telecom-paristech.fr. This
paper was presented in part in the IEEE Information Theory workshop, April 2015, Jerusalem, Israel.}
\thanks{This work is supported by the Australian Research Council
Discovery Project with ARC DP160101077.}}

\maketitle

\begin{abstract}
In this paper, we provide a solution to the open problem of computing the
Fourier transform of a binary function defined over $n$-bit vectors
taking $m$-bit vector values.
In particular, we introduce the two-modular Fourier transform (TMFT) of a binary function
$f:G\rightarrow {\cal R}$, where $G = (\mathbb{F}_2^n,+)$ is the group
of $n$ bit vectors with bitwise modulo two addition $+$, and ${\cal R}$ is a finite commutative ring of
characteristic $2$.
Using the specific group structure of $G$ and a sequence of nested subgroups of $G$,
we define the fast TMFT and its inverse.
Since the image ${\cal R}$ of the binary functions is a ring,
we can define the convolution between two functions $f:G\rightarrow {\cal R}$.
We then provide the TMFT properties, including the convolution theorem, which
can be used to efficiently compute convolutions.
Finally, we derive the complexity of the fast TMFT and the inverse fast TMFT.
\end{abstract}

\begin{IEEEkeywords}
Two-modular Fourier transform, binary functions, binary groups, group ring
\end{IEEEkeywords}

\IEEEpeerreviewmaketitle
\section{Introduction}

The Fourier transform is a fundamental tool in signal processing for spectral analysis
and is often used to transform a convolution between two real- or complex-valued functions into the
product of the respective transforms. In discrete-time signal processing, numerical evaluation of
the Fourier transform is based on the fast-Fourier transform (FFT), which enables to
efficiently compute convolutions \cite{Oppenheim10}.

More generally, the Fourier transforms of functions over finite Abelian groups $f: G \rightarrow \mathbb{C}$
(complex field) or $f: G \rightarrow \mathbb{Z}$ (ring of integers) have been extensively studied
\cite{Terras99}. For complex valued functions, when the group $G$ is cyclic, the Fourier transform
is the well-known discrete Fourier transform \cite{Terras99}. For complex valued functions, and
when $G$ is the additive group of $\mathbb{F}_2^{n}$, where $\mathbb{F}_2$ is the binary field, the Fourier transform
is provided by the well known Hadamard transform, commonly used for analyzing Boolean functions \cite{bolean}.
In computer science, harmonic analysis of Boolean functions is a powerful tool, which is used in the
theory of computational complexity (cf. the PCP Theorem in \cite[Chap. 22]{Arora09}).

The Fourier
transform of $f: G \rightarrow \mathbb{C}$, when
$G$ is finite and non-Abelian, is based on the complex
matrix representations of the non-Abelian group \cite{Terras99}. This Fourier transform
satisfies the convolution theorem,
which converts time-domain convolutions between functions into the product of the corresponding transforms.

The concept of Fourier transform was also extended to functions $f: G \rightarrow K$ defined over
finite group $G$ taking values in a finite field $K$, except for the case where the characteristic of
the field {\em divides} the order of the group. In general, for $f: G \rightarrow \mathcal{R}$,
where $G$ is an arbitrary group and $\mathcal{R}$ is a ring of prime characteristic $p$ co-prime
with the order of $G$, its Fourier transform is called the {\em $p$-modular Fourier transform},
which is similar to that of $f: G \rightarrow \mathbb{C}$, when $G$ is non-Abelian, but uses finite
field matrix representations of $G$ \cite{Terras99}.

An application of the $p$-modular Fourier transform, when $G$ is Abelian, enables to describe
Reed-Solomon codes and their decoding algorithms by a frequency domain interpretation \cite{Blahut84}.
In Reed-Solomon codes, ${\cal R}$ is the finite field $\mathbb{F}_{2^n}$ and the Abelian group $G$ is the cyclic
multiplicative group of $\mathbb{F}_{2^n}$. In this case, {\color{black} the order} of $G$ is $2^n-1$, which is {\em not divisible}
by the characteristic $2$ of the field. However, when the order of the group is {\em divisible} by
the characteristic $p$, and especially in the case of $p=2$ and $|G|=2^n$ (the order of $G$), the Fourier
transform has never been defined before.

In this paper, we provide a solution to this problem by introducing the {\em two-modular Fourier
transform} (TMFT) of a binary function
$f:G\rightarrow {\cal R}$, where $G = (\mathbb{F}_2^n,+)$ is the group
of $n$ bit vectors with bitwise modulo two addition $+$ and
${\cal R}$ is a finite commutative ring of characteristic $2$.
Furthermore, using the specific group structure of $G$ and a sequence of nested subgroups,
we introduce the fast TMFT and its inverse TMFT (ITMFT).

The TMFT is based on the two-modular two-dimensional representations of the additive group of
$\mathbb{F}_2$ and defines $n+1$ ``spectral components" as matrices over ``0'' and ``1'' in ${\cal R}$ of size
$2^k \times 2^k$, for $k=0 \ldots, n$. To develop ITMFT, we introduce a new operator
which extracts the top right corner
element of these matrices, since the trace operator used in the traditional Fourier transform
is not valid when the characteristic of the ring
${\cal R}$, $p=2$, divides the order of the group $|G|=2^n$.

{\color{black} When the ring ${\cal R}={\mathbb F}_2 = \{0,1\}$, the Hadamard transform
for $f: G \rightarrow \mathbb{C}$ can be used for faster convolution computations, since we can map
${\mathbb F}_2$ to $\{+1,-1\} \subset {\mathbb C}$ by $y=2x-1$.
However, if there is no such map from ${\cal R}$ to ${\mathbb C}$
then the traditional Fourier transform
for $f: G \rightarrow \mathbb{C}$ cannot be used for computing convolutions of $f: G \rightarrow {\cal R}$ functions.}
With our TMFT, we can provide the convolution theorem, since the TMFT preserves the multiplicative structure of the
ring ${\cal R}$, and enables efficient computations of multiplications
in the group ring ${\cal R}[G]$ \cite{Passman1977} of functions $f: G \rightarrow {\cal R}$.
Finally, we discuss the implementation and complexity of the fast TMFT and its inverse.


We expect the TMFT to have broad applications to problems in coding theory and computer science,
for example, in reliable computation of binary functions, network coding, {\color{black} cryptography}, and
classification of binary functions \cite{Carlet2010}.

The rest of this paper is organized as follows. Section~\ref{secII} reviews the classical
concept of Fourier transforms of functions defined over additive groups taking values in
complex or finite fields. In Section~\ref{secTMFT}, we present TMFT and {\em fast} TMFT of a binary function
$f:G\rightarrow {\cal R}$ defined over a finite commutative ring ${\cal R}$ of
characteristic $2$. In Section~\ref{secITMFT} we present the corresponding ITMFTs,
and in Section~\ref{secConv}, we prove the convolution theorem. In Section~\ref{complexity},
we discuss the implementation aspects and complexity of the proposed TMFT and ITMFT.
%
\section{Background}\label{secII}
In this section, we review the classical concept of Fourier transforms of functions defined over
additive groups taking values in complex or finite fields.
We highlight the essential mathematical ideas that are later used to define the TMFT.
In the following we assume the reader is familiar with the basic notions of {\em group, subgroup, quotient group,
homomorphism, the fundamental homomorphism theorem}, ring, and {\em field} \cite{Snaith2003}.

\begin{table*}[t]
\begin{center} \caption{\label{DFTexample} DFT example with $G=(\mathbb{Z}_6,+)$.}
\setlength{\unitlength}{0.6mm}
\begin{picture}(180, 121)(0,5)
\put(-60,-10){\includegraphics[width=8.5cm]{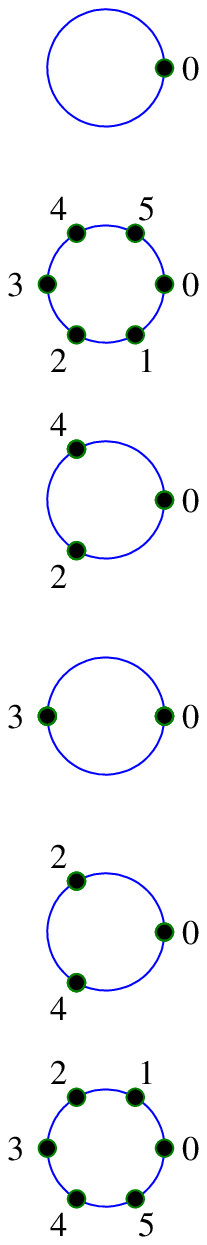} }
\put(-10,120){\line(1,0){210}} \put(0,0){\line(0,1){130}} \put(136,0){\line(0,1){130}}  \put(200,0){\line(0,1){130}}
\put(-7,123){$k$}
\put(2,123){$S_{k}\!=\!\mbox{Im}(\rho_k)\!=\! \{\rho_k(g), g\!\in\! G\!=\!\{0,1,2,3,4,5\} \}$ }
\put(140,123){$\mbox{Ker}(\rho_k), ~G/\mbox{Ker}(\rho_k)$}
\put(30,107){$\{1\}$}
\put(30,88){$\{1,e^{-\jmath\frac{2\pi}{6}},e^{-\jmath\frac{4\pi}{6}},
               e^{-\jmath\frac{6\pi}{6}},e^{-\jmath\frac{8\pi}{6}},e^{-\jmath\frac{10\pi}{6}}\}$}
\put(30,69){$\{1,e^{-\jmath\frac{4\pi}{6}},e^{-\jmath\frac{8\pi}{6}}\}$}
\put(30,50){$\{1,-1\}$}
\put(30,31){$\{1,e^{\jmath\frac{4\pi}{6}},e^{\jmath\frac{8\pi}{6}} \}$}
\put(30,12){$\{1,e^{\jmath\frac{2\pi}{6}},e^{\jmath\frac{4\pi}{6}},e^{\jmath\frac{6\pi}{6}},
               e^{\jmath\frac{8\pi}{6}},e^{\jmath\frac{10\pi}{6}}\}$}
\put(-7,107){$0$}
\put(-7,88){$1$}
\put(-7,69){$2$}
\put(-7,50){$3$}
\put(-7,31){$4$}
\put(-7,12){$5$}
\put(137,107){$~\{ 0,1,2,3,4,5 \},\{ 0 \}$}
\put(137,88){$~\{ 0 \},\{ 0,1,2,3,4,5 \}$}
\put(137,69){$~\{ 0,3 \},\{ 0,2,4 \}$}
\put(137,50){$~\{ 0,2,4 \},\{ 0,3 \}$}
\put(137,31){$~\{ 0,3 \},\{ 0,2,4 \}$}
\put(137,12){$~\{ 0 \},\{ 0,1,2,3,4,5 \}$}
\end{picture}
\end{center}
\end{table*}

\subsection{Algebraic view of the discrete Fourier transform} 

The discrete Fourier transform (DFT) is defined for $N$ samples of a real (or complex) discrete
time function $f: \mathbb{Z}_N \rightarrow \mathbb{C}$, where $\mathbb{Z}_N=\{0,1,\ldots, N-1\}$ is the time axis.
We can think of $f$ as discrete-time periodic function by $N$ samples. The DFT provides the well
known discrete spectrum of such function.
We observe that the time axis $\mathbb{Z}_N$ has an additive group structure
given by $G \triangleq (\mathbb{Z}_N,+)$ with addition mod $N$.
Hence we can think of $f: G \rightarrow \mathbb{C}$ as a complex valued function over the Abelian
group $G$.

Let the vector ${\bf f}= \left(f[n]\right)_{n=0}^{N-1}$ contain the $N$ values of the {\em time-domain}
function $f:G\rightarrow \mathbb{C}$. Then the DFT of ${\bf f}$ is given by the {\em frequency-domain}
vector ${\bf \hat{f}}= (\hat{f}[k])_{k=0}^{N-1}$, where
\begin{equation} \label{DFTdef}
\hat{f}[k] = \sum_{n=0}^{N-1} f[n]e^{-\jmath 2\pi\frac{nk}{N}},~~~~ k=0, \ldots, N-1,
\end{equation}
represents the transform of $f$ as a function  $\hat{f}:G\rightarrow \mathbb{C}$.
The corresponding frequency index $k$ also ranges in $\mathbb{Z}_N$ and the frequency axis has the same
group {\color{black} structure as $G$}. The inverse discrete Fourier transform (IDFT) of ${\bf \hat{f}}$ is given by
\begin{equation}
f[n] = \frac{1}{N}\sum_{k=0}^{N-1} \hat{f}[k]e^{\jmath 2\pi \frac{nk}{N}},~~~~ k=0, \ldots, N-1.
\end{equation}

The well known {\em DFT matrix} ${\bf F} = \{ e^{-\jmath 2\pi \frac{nk}{N}} \}_{n,k=0}^{N-1}$
is a unitary matrix such that
\begin{equation}\label{DFTmatrix}
{\bf \hat{f}}^T= {\bf F}{\bf f}^T ~~~~\mbox{and} ~~~~ {\bf f}^T= \frac{1}{N}{\bf F}^H{\bf \hat{f}}^T
\end{equation}
where $(\cdot)^T$ and $(\cdot)^H$ denote transposition and Hermitian transposition of a matrix,
respectively. The vectors ${\bf f}$ and ${\bf \hat{f}}$
are two `descriptions' of the signal $f[n]$ in different coordinate systems, namely the {\em
time basis} and the {\em frequency basis}.

{\color{black} We show how the group structure of the time axis can provide more insight into the DFT operation
by using the notions of {\em group representations} and {\em characters} (see Appendix A for a brief review).}

For the cyclic group $G=(\mathbb{Z}_{N},+)$, the scalar representation $\rho_k$ is the homomorphism from $G$ to
the unit circle in the complex plane ${\mathcal S}=\{ z\in \mathbb{C}: |z|=1\}$, given by
\[
\rho_k: G\rightarrow S_{k} \subset {\mathcal S} ~~~~\rho_k(n)\triangleq e^{-\jmath2\pi \frac{nk}{N}}
\]
for $k=0,\ldots, N-1$, and the image of $\rho_k$ is the set of  distinct points on the unit circle
\[
S_{k}\triangleq \mbox{Im}(\rho_k) = \left\{1,e^{-\jmath2\pi\frac{k}{N}}, e^{-\jmath2\pi\frac{2k}{N}},  \cdots, e^{-\jmath2\pi\frac{(N-1)k}{N}} \right\}.
\]

The representation $\rho_k$ is a group homomorphism transforming $G$ {\color{black} into the group of complex roots of unity}
$S_{k}$, i.e., for any $g_1,g_2\in G$
\[
\rho_k(g_1 + g_2) = \rho_k(g_1) \rho_k(g_2)
\]
since
\[
 e^{-\jmath2\pi \frac{(g_1 + g_2)k}{N}}= e^{-\jmath2\pi \frac{g_1 k}{N}} e^{-\jmath 2\pi \frac{g_2 k}{N}}.
 \]

We now illustrate the relation between the DFT and the representation of a cyclic group using the
example below with $G=(\mathbb{Z}_6, +)$.
\begin{myexample}
Using Definition \ref{def:ineqrep} (see Appendix A) in the scalar case, Table \ref{DFTexample} illustrates all the inequivalent scalar representations
$\rho_k:G\rightarrow {\mathcal S}_k$ for $k=0,\dots,5$.
Some representations are faithful (e.g., $\rho_1$ and $\rho_5$), the others are not. According to
the {\em fundamental homomorphism theorem} of groups \cite[Th.~1.5.6]{Snaith2003}, the image $S_k$ is isomorphic to
the quotient group $G/{\mbox{Ker}(\rho_k)}$, where $\mbox{Ker}(\rho_k)$ is a normal subgroup of $G$.
$\hfill \square$
\end{myexample}

We can formally rewrite the DFT in (\ref{DFTdef}) as
\begin{equation} \label{FormalDFT}
\hat{f} [k] = \sum_{g\in G} f[g] \rho_k(g)  ~~~ k=0, \ldots, N-1
\end{equation}
Let $(\cdot)^*$ denote complex conjugation. Then we observe that the {\color{black} pairwise orthogonal complex vectors ${\boldsymbol\psi}_k = [\rho_k^{*}(g)]_{g\in G}$},
form the discrete Fourier basis vectors  in ${\mathbb C}^N$ (i.e., the columns of the DFT matrix ${\bf F}$
in (\ref{DFTmatrix})). This is shown in Table \ref{DFTmatZ6} for  $G=(\mathbb{Z}_6, +)$.
The formal DFT in (\ref{FormalDFT}) can also be interpreted as the {\color{black} complex} scalar product
\begin{equation}
\hat{f} [k] =  \langle{\bf f}, {\boldsymbol\psi}_k \rangle ~~~ k=0, \ldots, N-1
\end{equation}
which gives the projection of the time domain vector ${\bf f}$ along the Fourier basis vector
${\boldsymbol\psi}_k$. 

{\color{black} We now discuss how the fast Fourier transform (FFT) naturally stems from the group structure of $G=({\mathbb Z}_N,+)$.
From the fundamental homomorphism theorem \cite[Th.~1.5.6]{Snaith2003}, since $\mbox{Ker}(\rho_k)$ is a subgroup of $G$, the direct product of
$\mbox{Ker}(\rho_k)$ and $G/{\mbox{Ker}(\rho_k)}$ is isomorphic to $G$, i.e.,}
\begin{eqnarray}
G &=& \{g = u+v | u\in \mbox{Ker}(\rho_k), v\in G/\mbox{Ker}(\rho_k) \} \nonumber\\
&\cong& \mbox{Ker}(\rho_k) \times G/\mbox{Ker}(\rho_k),
\end{eqnarray}
and
\begin{equation}\label{eq:kernelFFT}
\rho_k(u)=\rho_k(0)=1 \in S_k   ~~~{\mbox{for~all}}~~u\in \mbox{Ker}(\rho_k)~.
\end{equation}

Then we can compute the DFT more efficiently as
\begin{eqnarray}\label{FFT}
\hat{f}[k] &=& \sum_{g\in G} f[g] \rho_k(g)\nonumber\\
&=&  \sum_{v\in G/\mbox{\scriptsize Ker}(\rho_k)}\sum_{u\in \mbox{\scriptsize Ker}(\rho_k)} f[u+v] \rho_k(u+v) \nonumber\\
&=&  \sum_{v\in G/\mbox{\scriptsize Ker}(\rho_k)} \left(\; \sum_{u\in \mbox{\scriptsize Ker}(\rho_k)} f[u+v]\rho_k(u)\right) \rho_k(v)\nonumber\\
&=&  \sum_{v\in G/\mbox{\scriptsize Ker}(\rho_k)} \left(\; \sum_{u\in \mbox{\scriptsize Ker}(\rho_k)} f[u+v]\right) \rho_k(v)
\end{eqnarray}
for $k=0, \ldots, N-1$.
The last equality in (\ref{FFT}) is due to (\ref{eq:kernelFFT}).
From (\ref{FFT}), we observe how the number of multiplications reduces from $|G|^2 = N^2$ to
\[
\sum_{k} |G/\mbox{Ker}(\rho_k)|~.
\]
In the above example, the number of multiplications reduces from $36$ to $21$. Note that by taking advantage
of the Hermitian symmetry of the DFT matrix ${\bf F}$, the number of multiplications can be further reduced to $12$.


\begin{table}
\begin{center}\caption{\label{DFTmatZ6} Fourier basis vectors DFT example with $G=(\mathbb{Z}_6,+)$.}
\begin{tabular} {|c||cccccc||c|}
\hline
$g\in G$                 & $0$ & $1$ & $2$ & $3$ & $4$ & $5$ & ~ \\ \hline\hline
$ \rho_0^*(g)$ & $1$ & $1$ & $1$ & $1$ & $1$ & $1$ & \!${\boldsymbol\psi}_0$\! \\
$ \rho_1^*(g)$ & $1$ & $e^{+\jmath\frac{2\pi}{6}}$ & $e^{+\jmath\frac{4\pi}{6}}$
& $e^{+\jmath\frac{6\pi}{6}}$ & $e^{+\jmath\frac{8\pi}{6}}$ & $e^{+\jmath\frac{10\pi}{6}}$
 & \!${\boldsymbol\psi}_1$\! \\
$\rho_2^*(g)$ & $1$ & $e^{+\jmath\frac{4\pi}{6}}$ & $e^{+\jmath\frac{8\pi}{6}}$
                     & $1$ & $e^{+\jmath\frac{4\pi}{6}}$ & $e^{+\jmath\frac{8\pi}{6}}$
 & \!${\boldsymbol\psi}_2$\! \\
$\rho_3^*(g)$ & $1$ & $-1$ & $1$ & $-1$ & $1$ & $-1$  & \!${\boldsymbol\psi}_3$\! \\
$ \rho_4^*(g)$ & $1$ & $e^{-\jmath\frac{4\pi}{6}}$ & $e^{-\jmath\frac{8\pi}{6}}$
                     & $1$ & $e^{-\jmath\frac{4\pi}{6}}$ & $e^{-\jmath\frac{8\pi}{6}}$
 & \!${\boldsymbol\psi}_4$\! \\
$\rho_5^*(g)$ & $1$ & $e^{-\jmath\frac{2\pi}{6}}$ & $e^{-\jmath\frac{4\pi}{6}}$
& $e^{-\jmath\frac{6\pi}{6}}$ & $e^{-\jmath\frac{8\pi}{6}}$ & $e^{-\jmath\frac{10\pi}{6}}$
 & \!${\boldsymbol\psi}_5$\! \\ \hline
\end{tabular}
\end{center}
\end{table}
\subsection{Fourier Transform of $f: G \rightarrow \mathbb{C}$ for {\color{black} arbitrary} $G$}
The classical notion of Fourier transform over arbitrary finite groups is based on the $n$-dimensional
representations of group elements by complex $n\times n$ matrices in $GL(n,\mathbb C)$ (see Appendix A). It generalizes
the well known discrete Fourier transform, which is naturally defined over a cyclic group
(additive Abelian group). In the general case where $G$ is non-Abelian, the group element
representations are matrices and we have
\begin{mydefinition}  \label{def_groupFT} (\cite{Terras99})
Given a finite group $G$, the Fourier transform of a function
$f:G\rightarrow {\mathbb C}$ evaluated for a given representation
$\rho: G \rightarrow GL({d_\rho},{\mathbb C})$ of  $G$, of dimension $d_\rho$, is given by
the $d_\rho \times d_\rho$ matrix
\[
\hat{f}(\rho) = \sum_{g\in G} f(g) \rho(g).
\]
The complete Fourier transform is obtained by considering all the $\rho$'s in the set $\{\rho_k\}$
of all inequivalent irreducible representations of $G$ (see Appendix A).
$\hfill \square$
\end{mydefinition}

\begin{mydefinition}(\cite{Terras99})\label{def_groupIDFT}
The inverse Fourier transform evaluated at $g \in G$ is given by
\begin{equation}\label{eq:IDFTDef6}
f(g) = \frac{1}{|G|} \sum_k d_{\rho_k} \,\mbox{Tr} \left(\rho_k(g^{-1}) \hat{f}(\rho_k)  \right)
\end{equation}
where $|G|$ is the order of the group $G$.
$\hfill \square$
\end{mydefinition}

{\color{black} Note that Definitions \ref{def_groupFT} and \ref{def_groupIDFT} generalize the DFT/IDFT for the Abelian group $G=({\mathbb Z}_N,+)$.}
The above Fourier transform is well defined for complex valued functions over finite groups $G$
and can be used to transform convolution in the `time-domain' defined
as\footnote{We adopt the conventional multiplicative group notation for non-Abelian groups.} (\cite{Terras99})
\begin{equation}\label{EQconv}
(f_1\ast f_2)(g) = \sum_{h\in G} f_1(h^{-1}g)f_2(h)~~\mbox{for~all}~~g\in G
\end{equation}
into the product of the `frequency domain' transforms, i.e., \cite{Terras99}
\[
\widehat{(f_1\ast f_2)}(\rho) = \hat{f_1}(\rho) \hat{f_2}(\rho)~.
\]

\subsection{Fourier Transform of $f: G \rightarrow K$}
We now consider the Fourier transform of functions over a finite group $G$ taking
values in a finite field $K={\mathbb F}_{p^n}$ of prime characteristic $p$, where $n$ is positive integer.
Let $\alpha$ be a primitive element of $K$ \cite{Pollard1971}, then we can list all the elements
in ${\mathbb F}_{p^n}$ as $\{0, 1,\alpha,\ldots,\alpha^{p^n-2}\}$.
\begin{mydefinition}\label{def:finitefieldFT}
(\cite{Pollard1971}) For an Abelian group $G=({\mathbb Z}_N,+)$, where $N$ is a divisor of $p^n -1$
and $p$ is coprime with $N$, we define the {\em finite field Fourier transform} of
$f: {\mathbb Z}_N\rightarrow {\mathbb F}_{p^n}$ as
\[
\hat{f}[k] = \sum_{n=0}^{N-1} f[n] \alpha^{nk}
\]
and its {\em finite field inverse Fourier transform} as
\[
f[n] = \frac{1}{N}\sum_{k=0}^{N-1} \hat{f}[k] \alpha^{-nk}.
\]
~$\hfill\square$
\end{mydefinition}
The finite field inverse Fourier transform exists only if $p$ is co-prime with $N=|G|$.
This definition can be reformulated as in (\ref{FormalDFT}) using the
scalar representations $\rho_k : G\rightarrow K$ that are defined by $N$ vectors
\[
[1, \alpha^k,\alpha^{2k},\ldots, \alpha^{(N-1)k}] ~~~~~~ \mbox{for}~~k=0,\ldots, N-1.
 \]

We note that this Fourier transform is only defined when $G=({\mathbb Z}_N,+)$ is
isomorphic to a subgroup of the cyclic multiplicative group of ${\mathbb F}_{p^n}$.
For any other non-Abelian group $G$, we need to modify Definition \ref{def_groupFT} by replacing the group representations with
the $p$-{\em modular representations} of $G$ defined below. 

\begin{mydefinition}\label{def:p-modularFT}
A $p$-{\em modular representation} of a group $G$ over a field $K$ of prime characteristic $p$
is a group homomorphism $\pi: G \mapsto GL(n,K)$, such that
the binary operation of two group elements corresponds to the matrix multiplication.~$\hfill\square$
\end{mydefinition}

\section{The two-modular Fourier transform of binary functions}\label{secTMFT}
%
We now focus on binary functions (i.e., from $n$ bit vectors to $m$ bit vectors)
$f: G\rightarrow {\cal R}$ where $G=(\mathbb{F}_2^n, +)$ is the group of $n$-bit binary
vectors with bitwise mod two addition $+$, and ${\cal R}$ is a finite commutative
ring of characteristic 2. For example, we can choose ${\cal R}=(\mathbb{F}_{2}^m, +, \wedge)$,
where addition and multiplication are defined by bitwise $+$ (XOR) and  $\wedge$ (AND) binary logic
operators, respectively. Another example is a polynomial ring $\mathbb{F}_2[X]/\phi(X)$, where $\phi(X)$
is an arbitrary polynomial of degree $m$. In the special case where $\phi(X)$ is an irreducible polynomial,
${\cal R}$ is the finite field $K=\mathbb{F}_{2^m}$.
The elements of ${\cal R}$ can be represented as binary coefficient polynomials of degree less than $m$,
where the ring operations are polynomial addition and multiplication mod $\phi(X)$.
In the following, we will denote the zero and one elements of the ring ${\cal R}$ as $0$ and $1$, and $1+1=0 \in {\cal R}$.
In the special case of ${\cal R}=(\mathbb{F}_{2}^m, +, \wedge)$, we have $0\rightarrow {\bf 0}_m$ and $1\rightarrow {\bf 1}_m$,
where ${\bf 0}_m$ and ${\bf 1}_m$ denote the $m$-bit all-zero and all-one vectors.
Nevertheless, we use $0$ and $1$ in all cases for simplicity of notation.

For convenience of notation, we will label the $n$-bit vectors
${\bf b}=(b_1,\ldots,b_n) \in G$ using the corresponding decimal values $\{0,\ldots, 2^{n}-1\}$, i.e.,
{\color{black}
\begin{equation} \label{eq:Drepres}
D({\bf b}) = \sum_{k=1}^n b_k 2^{n-k}
\end{equation}
and its inverse as
\begin{equation} \label{eq:invDrepres}
{\bf b}\triangleq D^{-1}(j)
\end{equation}}
for any decimal $j\in \{0,\ldots, 2^{n}-1\}$. In the following, we will first introduce the
{\em two-modular representations} of binary groups. Then we will introduce
the TMFT and the fast TMFT.

\subsection{Two-modular representations of binary groups}
\begin{mydefinition}
The {\em two-modular representation} of the binary group $C_2=(\mathbb{F}_2,+)=\{0,1\}$ is
defined as $2\times 2$ matrices over $\{0,1\}\in{\cal R}$, i.e., $\pi_1(C_2)=\{E_0,E_1\}$, where
\[
\pi_1(0) =  E_0 \triangleq\left(
  \begin{array}{cc}
    1 & 0 \\
    0 & 1 \\
  \end{array}
\right)
 ~~\mbox{and}~~
\pi_1(1) =  E_1 \triangleq\left(
  \begin{array}{cc}
    1 & 1 \\
    0 & 1 \\
  \end{array}
\right).
\]
$\hfill\square$
\end{mydefinition}
\begin{mylemma}\label{lemma:ematrix}
The $n$-fold direct product group  $C_2^n=(\mathbb{F}_2^n,+)$ can be faithfully represented as the
Kronecker product of the representations of $C_2$, i.e.,
\begin{eqnarray}
\pi_n(C_2^n) &\triangleq & \pi_1(C_2)\otimes \cdots \otimes \pi_1(C_2)\nonumber\\
&=& \pi_{n-1}(C_2^{n-1}) \otimes \pi_1(C_2). \label{repres_Kronecker}
\end{eqnarray}
Specifically, the matrix representation of a group element ${\bf b}=(b_1,\ldots,b_n)$  is,
\begin{equation} \label{eq:def_Eb}
E_{\bf b} \triangleq \pi_1(b_1)\otimes \cdots \otimes \pi_1(b_n).
\end{equation}
$\hfill\square$
\end{mylemma}

{\em Proof}:
We need to show that $\pi_n$ is an injective homomorphism.
For $n=0$ and 1, it is straightforward. For $n\geq 2$, we prove it by induction using the
recursion (\ref{repres_Kronecker}). Thus it is enough to consider the case $n=2$ and show
that $\pi_2$ is a group homomorphism between $C_2^2$ and $\pi_2(C_2^2)$,
i.e., that
\[
\pi_2(b_1+ c_1,b_2+ c_2)=\pi_2(b_1,b_2) \cdot \pi_2(c_1,c_2),
\]
or equivalently,
{\color{black} \[
E_{(b_1+ c_1,b_2+ c_2)}=E_{(b_1,b_2)} \cdot E_{(c_1,c_2)}~.
\] }
From (\ref{repres_Kronecker}) we have $\pi_2=\pi_1\otimes\pi_1$ then
\begin{eqnarray}  \label{eq_Lemma1}
E_{(b_1,b_2)} \cdot E_{(c_1,c_2)} &=& (E_{b_1}\otimes E_{b_2}) \cdot (E_{c_1}\otimes E_{c_2}) \nonumber\\
&=& (E_{b_1}E_{c_1}) \otimes (E_{b_2}E_{c_2}) \\
&=& E_{(b_1+ c_1,b_2+ c_2)}~.\nonumber
\end{eqnarray}
Then, we note that $E_{\bf b} ={\bf I}_{2^n}$ holds only for  ${\bf b}={\bf 0}_n$. This proves the homomorphism is injective, since the kernel of $\pi_n$
is only the all zero binary vector.
$\hfill\square$

Finally we define the representation of the trivial group $\{0\}$ as
$\pi_0(\{0\})\triangleq 1 \in {\cal R}$.

\begin{myexample}
The two-modular representation of $G=\mathbb{F}_2^2$ is given by:
\[ \footnotesize
E_{00}\!=\!\left(\!\!
         \begin{array}{c}
           1~0~0~0 \\
           0~1~0~0 \\
           0~0~1~0 \\
           0~0~0~1
         \end{array}\!\!
       \right) ~~
E_{01}\!=\!\left(\!\!
         \begin{array}{c}
           1~1~0~0 \\
           0~1~0~0 \\
           0~0~1~1 \\
           0~0~0~1
            \end{array}\!\!
       \right) ~~\]
 \[ \footnotesize
E_{10}\!=\!\left(\!\!
         \begin{array}{c}
           1~0~1~0 \\
           0~1~0~1 \\
           0~0~1~0 \\
           0~0~0~1
            \end{array}\!\!
       \right) ~~
E_{11}\!=\!\left(\!\!
            \begin{array}{c}
           1~1~1~1 \\
           0~1~0~1 \\
           0~0~1~1 \\
           0~0~0~1
            \end{array}\!\!
       \right)~.
\]
$\hfill\square$
\end{myexample}
We list below a few simple properties of the representation matrices.
\begin{myproperty} \label{prop:p1}
The $2\times 2$ matrices $E_0$ and $E_1$ are upper triangular and have antidiagonal symmetry.
Hence, any $E_{\bf b}$ defined by the Kroneker product in (\ref{eq:def_Eb}) is also upper
triangular and antidiagonal symmetric. $\hfill\square$
\end{myproperty}
\begin{myproperty} \label{prop:p2}
Any linear combination of $E_{\bf b}$ matrices is
upper triangular and antidiagonal symmetric. $\hfill\square$
\end{myproperty}
\begin{myproperty} \label{prop:p3}
From Lemma \ref{lemma:ematrix}, we have
$E_{\bf b} ={\bf I}_{2^n}$ only for  ${\bf b}={\bf 0}_n$, where ${\bf I}_{2^n}$ denotes the
${2^n} \times {2^n}$ identity matrix. $\hfill\square$
\end{myproperty}
\begin{myproperty} \label{prop:p4}
The top right corner element of $E_{\bf b}$ is $1$ only for ${\bf b}={\bf 1}_n$.  $\hfill\square$
\end{myproperty}
\begin{myproperty} \label{prop:p5}
{\color{black} The main diagonal of $E_{\bf b}$'s is all $1$s}, and hence the top left corner element is always $1$.
\end{myproperty}

\subsection{The TMFT and the fast TMFT}
Let us consider a sequence of nested subgroups $H_k \cong C_2^k$ of $G=C_2^n$, namely
\begin{equation}\label{eq:nestedsubG}
H_0=\{{\bf 0}_n\} \lhd H_1 \lhd \cdots \lhd H_k \lhd \cdots \lhd H_{n-1} \lhd G ~
\end{equation}
where $\{{\bf 0}_n\}$ denotes the trivial group with only one element, the $n$-bit zero vector.

There are many possible choices for such sequence $H_k\cong C_2^k$ and
we now choose a specific one, which results in a simpler notation.
In particular, we choose $H_k$, $k=1,\ldots,n$,  to be the set of $n$-bit
vectors with the first $n-k$ bits set to zero, i.e.,
\begin{equation}\label{eq:Hk}
H_k = \{ (0,\ldots, 0, b_{n-k+1}, \ldots, b_n) | b_i\in \{0,1\} \}\cong C_2^k,
\end{equation}
and $k=1,\ldots,n$.
Then we consider the quotient groups $G/H_k$, which are the sets of $n$-bit
vectors with the last $k$ bits set to zero, i.e.,
\begin{equation}\label{eq:gHk}
G/H_k = \left\{ \begin{array}{l}
\{(b_1, \ldots, b_{n-k}, 0,\ldots, 0) | b_i\in \{0,1\} \}\\
\hspace{3.75cm}k=1,\ldots n-1 \\
\{{\bf 0}_n \} \hspace{3cm}
k=n~
\end{array} \right.
\end{equation}
\begin{table}
\caption{Nested subgroups and corresponding quotient groups of $C_2^3$ \label{Fig_tables}}
\[
\begin{tabular}{|c|c|}
              \hline
              $D(g)$& $G$\\
              \hline
              0 & 000 \\
              1 & 001 \\
              2 & 010 \\
              3 & 011 \\
              4 & 101 \\
              5 & 101 \\
              6 & 110 \\
              7 & 111 \\
              \hline
      \end{tabular} ~~
\begin{tabular}{|c|c|}
\hline
              & $H_1$\\
              \hline
  0 & 000 \\
  1 & 001 \\
  \hline
  \hline
              & $G/H_1$\\
              \hline
  0 & 000 \\
  2 & 010 \\
  4 & 100 \\
  6 & 110 \\
  \hline
\end{tabular} ~~
\begin{tabular}{|c|c|}
\hline
              & $H_2$\\
              \hline
  0 & 000 \\
  1 & 001 \\
  2 & 010 \\
  3 & 011 \\
  \hline
  \hline
              & $G/H_2$\\
              \hline
  0 & 000 \\
  4 & 100 \\
  \hline
\end{tabular}
\]
\end{table}
Table~\ref{Fig_tables} shows an example of $G$, $H_1$, $H_2$, $G/H_1$, and $G/H_2$ for $n=3$ bits, where we index each element using its corresponding decimal value from (\ref{eq:Drepres}).
Let $d_k \in H_k$ be the $n$-bit all-zero vector except for its $(n-k+1)$-th bit set to $1$, i.e. {\color{black}
\[
d_k = (\underbrace{0,\ldots,0}_{n-k}, \underbrace{1}_{n-k+1}, \underbrace{0, \ldots,0}_{k-1})
\]}
Let us consider the binary subgroups of $H_k$
generated by $d_k$, i.e., {\color{black} $\langle d_k \rangle = H_k/H_{k-1} = \{{\bf 0},d_k\}\cong C_2$ and  $H_k/\langle d_k \rangle = H_{k-1}$}.
Then we have the following decomposition
\begin{equation}  \label{eq:tree_decomp}
\underbrace{G}_{2^n} = \underbrace{H_k/\langle d_k\rangle}_{2^{k-1}}  \times  \underbrace{\langle d_k \rangle}_{2} \times \underbrace{G/H_k}_{2^{n-k}}  ~~~k=1,\ldots, n
\end{equation}
where cardinalities of the component subgroups are indicated below each one and
\begin{equation}\label{eq:Hkdk}
H_k/\langle d_k\rangle = \left\{ \begin{array}{l}
\{{\bf 0}_n \}  \hspace{3.5cm}  k=1 \\
\{ (0,\ldots, 0, b_{n-k+2}, \ldots, b_n) | \\
\hspace{1cm}b_i\in \{0,1\} \}\cong C_2^{k-1} ~~ k=2,\ldots,n~
\end{array} \right.
\end{equation}

{\color{black} For any $g=(b_1,\ldots, b_n)\in G$, we have $g=u+v$ or $g=u+v+d_k$,
where $u=(0,\ldots, 0, b_{n-k+2}, \ldots, b_n) \in H_k/\langle d_k\rangle$ for $k=2,\ldots,n$ (or $u\in \{{\bf 0}_n\}$ for $k=1$),
and $v=(b_1, \ldots, b_{n-k}, 0,\ldots, 0)\in G/H_k$ for $k=1,\ldots,n-1$ (or $v\in \{{\bf 0}_n\}$ for $k=n$). The element $d_k$ is the $n$-bit all-zero vector except for its $(n-k+1)$-th bit set to $1$, as defined above.}

We now define $\sigma_k: H_k/\langle d_k\rangle \rightarrow C_2^k$ as a map converting the $n$
bit vectors in $H_k/\langle d_k\rangle$ to $k$ bit vectors in $C_2^k$, which removes the first $n-k$ zero bits of the $n$ bit vectors of {\color{black} $H_k/\langle d_k\rangle$}.
Specifically, for any $u\in H_k/\langle d_k\rangle$, we have
\begin{equation}\label{eq:sigma_u}
\sigma_k(u) \triangleq \left\{ \begin{array}{cl}
0  &  k=1. \\
(0, b_{n-k+2}, \ldots, b_n) & k=2,\ldots,n
\end{array} \right.
\end{equation}
We note that $\mbox{Im}(\sigma_k)$ does not contain any pair of complementary vectors. All the complementary vectors are in the $\mbox{Im}(\sigma_k)$, where
\begin{equation}\label{eq:sigma_u_bar}
\overline{\sigma_k(u)} \triangleq \left\{ \begin{array}{cl}
1  &  k=1. \\
(1, {\bar b}_{n-k+2}, \ldots, {\bar b}_n) & k=2,\ldots,n
\end{array} \right.
\end{equation}
where ${\bar b}_i$ represents the binary complement of $b_i\in \{0,1\}$.

\begin{mylemma}\label{lemma:sigma_k}
The map $\sigma_k$ is a homomorphism, i.e., given $u_1, u_2 \in H_k/\langle d_k\rangle$, we have
$\sigma_k (u_1 +u_2)=\sigma_k (u_1) + \sigma_k (u_2)$ and $\sigma_k ({\bf 0}_n)={\bf 0}_k$, but the map $\overline{\sigma_k}$ is not.
$\hfill\square$
\end{mylemma}
{\em Proof}: The proof is straightforward. ~~$\hfill\square$

Let $\tau_k: G \mapsto C_2^k$, $k=1,\ldots,n$, be a map with {\color{black} image}
$\mbox{Im}(\tau_k)=C_2^k$, which defines the $k$-bit vector index
${\bf b}=\tau_k(g)$ of $E_{\bf b}=E_{\tau_k(g)}$, for all $g \in G$.
In particular, for any $g \in G$,
$\tau_k (g)$ is defined as
\begin{equation}\label{eq:tau_k_u}
\tau_k (g) \triangleq \left\{ \begin{array}{cl}
\sigma_k (u) &  ~~{\mbox{if}}~~ g=u+v~ \mbox{for~some}~\\
&\hspace{0.2cm}~u\in H_k/\langle d_k\rangle ~\mbox{and}~v\in G/{H_k}  \\
\overline{\sigma_k (u)} & ~~{\mbox{if}}~~ g=u+v+d_k~\mbox{for~some} \\
&\hspace{0.2cm}~u\in H_k/\langle d_k\rangle ~\mbox{and}~v\in G/{H_k}
\end{array} \right.
\end{equation}

\begin{mylemma}\label{lemma:tau_kgrouphomo}
The map $\tau_k: G \mapsto C_2^k$ is a group homomorphism, i.e., $\tau_k (g+w)=\tau_k (g) + \tau_k (w)$, for $g,w\in G$,
and $\mbox{Ker} (\tau_k) = G/H_k$. $\hfill\square$
\end{mylemma}
{\em Proof}: {\color{black} The proof is given in Appendix B}. ~~$\hfill\square$

We now consider the two-modular representations $\pi_k(\tau_k (g)) = E_{\tau_k (g)}$ of $G$,
with {\color{black} image} $\mbox{Im}(\pi_k)=\pi_k(C_2^k)=\{E_{\tau_k (g)}: g\in G\}$ with $2^k$ elements isomorphic to
the nested subgroups $H_k$, i.e.,
\begin{eqnarray*}
H_0 & \cong & \mbox{Im}(\pi_0)  = \{1\} \\
H_1 & \cong & \mbox{Im}(\pi_1)  = \{E_0, E_1\}\\
H_2 & \cong & \mbox{Im}(\pi_2)  = \{E_{00}, E_{01}, E_{10},E_{11}\}\\
H_3 & \cong & \mbox{Im}(\pi_3)  = \{E_{000}, E_{001}, E_{010}, E_{011}, \\
&&\hspace{1.7cm}E_{100}, E_{101},  E_{110}, E_{111}\}\\
&\vdots &
\end{eqnarray*}
the Fourier basis `vectors' $\boldsymbol\psi_k = [E_{\tau_k (g)}: g\in G]$ are the $2^n$-component vectors (indexed by $g$) of $2^k\times 2^k$ matrices from
the set $\mbox{Im}(\pi_k)$. 

The projection of $f$ on the $k$-th Fourier basis vector $\boldsymbol\psi_k$, for $k=0,\ldots, n$, gives
the corresponding Fourier coefficient $\hat{f}_k$, which is a $2^k\times 2^k$ matrix.
\begin{mydefinition}\label{Def:absFT}({\bf TMFT}).
We define the $k$-th Fourier coefficients of the TMFT for $k=1,\ldots,n$ as the $2^k \times 2^k$ matrix
\begin{equation}\label{def:FT}
\hat{f}_k = \langle f, \boldsymbol\psi_k \rangle \triangleq \sum_{g\in G} f(g) E_{\tau_k(g)}
\end{equation}
where $E_{\tau_k(g)}$ is the $g$-th element of the vector $\boldsymbol\psi_k$
and for $k=0$ we define
\begin{equation}\label{def:FT0}
\hat{f}_0 = \langle f, \boldsymbol\psi_0 \rangle \triangleq \sum_{g\in G} f(g)
\end{equation}
and we refer to $\hat{f}_0$ as the `DC-component' of $f$.
 ~~~$\hfill\square$
\end{mydefinition}

We are now ready to define the {\em fast TMFT} to compute (\ref{def:FT}) more efficiently
by collecting the terms with the same $E_{\tau_k(g)}$.
{\color{black}
\begin{mylemma}\label{fastTMFT} ({\bf fast TMFT})
The $k$-th Fourier coefficients $\hat{f}_k$ of the {\em fast TMFT} for
$k=1,\ldots, n$ can be efficiently computed as
\begin{eqnarray}\label{eq_FT}
\hat{f}_k&=&\sum_{u\in H_k/\langle d_k \rangle} \left\{\left[\sum_{v\in G/H_k} f(u+v)\right] E_{\sigma_k(u)} \right.\nonumber\\
              && \hspace{5mm} +\left.\left[\sum_{v\in G/H_k}f(u+d_k+v)\right] E_{\overline{\sigma_k(u)}}\right\}~
\end{eqnarray}
 For $k=0$, (\ref{def:FT0}) holds as is. $\hfill\square$
\end{mylemma} }

{\color{black}
{\em Proof:}
We note that, in Definition \ref{Def:absFT}, for $g\in G$, there are $2^n$ matrices $E_{\tau_k(g)}$ of size $2^k\times 2^k$
in the computation of $\hat{f}_k$, $k=1,\ldots,n$. Among these matrices $E_{\tau_k(g)}$, there are $2^k$ distinct
ones in pairs of $E_{\sigma_k(u)}$ and $E_{\overline{\sigma_k(u)}}$, where $u\in H_k/\langle d_k\rangle$, according to (\ref{eq_FT}).
Hence, the fast TMFT can collect the $2^{n-k}$ terms with the same $E_{\tau_k(g)}$, leading to a reduced computation complexity (see details on complexity analysis in Section~\ref{complexity}).}

$\hfill\square$

\begin{myexample}\label{exampleFFT}
The Fourier coefficients of the fast TMFT for a function over $G=C_2^3$
can be computed using $H_1$ and $H_2$ defined in (\ref{eq:Hk}) as
\begin{eqnarray}
\hat{f}_0 \hspace{-3mm}&=&\hspace{-4mm}
  \sum_{g\in G} f(g) \label{Ex2FT0}\\
\hat{f}_1 \hspace{-3mm}&=&\hspace{-6mm}  \sum_{v\in G/H_1}  f(\underbrace{(000)}_{u}+v) E_{0} + f(\underbrace{(000)}_{u}+\underbrace{(001)}_{d_1}+v) E_{1} \label{Ex2FT1}\\
\hat{f}_2 \hspace{-3mm}&=&\hspace{-6mm}  \sum_{u\in H_2/\langle d_2 \rangle} \sum_{v\in G/H_2}\hspace{-4mm}  f(u\hspace{-1mm}+\hspace{-1mm}v) E_{\sigma_2(u)}\hspace{-1mm}
              +\hspace{-1mm}   f(u\hspace{-1mm}+\hspace{-1mm}(010)\hspace{-1mm}+\hspace{-1mm}v) E_{\overline{\sigma_2(u)}} \nonumber\\
              &&\label{Ex2FT2}\\
\hat{f}_3 \hspace{-3mm}&=&\hspace{-6mm}  \sum_{u\in H_3/\langle d_3 \rangle} f(u) E_{\sigma_3(u)} + f(u+(100)) E_{\overline{\sigma_3(u)}}~\label{Ex2FT3}
\end{eqnarray}

The sum indices in (\ref{Ex2FT0}), (\ref{Ex2FT1}), (\ref{Ex2FT2}), and (\ref{Ex2FT3})
are based upon these group elements listed in Table~\ref{Fig_tables}.
For example, in (\ref{Ex2FT2}), given ${\langle d_2 \rangle} = \{000, 010\}$,
according to (\ref{eq:Hkdk}) and (\ref{eq:sigma_u}), we choose $u\in H_2/\langle d_2 \rangle=\{000,001\}$,
and thus we obtain the corresponding $\sigma_2(u) \in \{00, 01\}$ with the associated matrices $\{ E_{00}, E_{01}\}$. Then $\overline{\sigma_2(u)} \in \{11, 10\}$ and
the associated matrices are $\{E_{11}, E_{10}\}$.

Similarly, in (\ref{Ex2FT3}), given ${\langle d_3 \rangle} = \{000, 100\}$,
we choose $u\in H_3/\langle d_3 \rangle=\{000,001,010,011\}$, which yields $\sigma_3(u) \in \{000, 001$ $010, 011\}$ with the associated
matrices $\{E_{000}, E_{001}$, $E_{010}, E_{011}\}$,
and $\overline{\sigma_3(u)} \in \{111, 110$, $101, 100\}$ with the associated matrices
$\{E_{111}, E_{110}, E_{101}, E_{100}\}$.

\begin{figure}[t*]
 \begin{center}
 \setlength{\unitlength}{0.6mm}
 \begin{picture}(120, 160)(-9,0)
 \put(-10,0){$u\in H_1/\langle d_1 \rangle, ~~~H_2/\langle d_2 \rangle,~~~~ H_3/\langle d_3 \rangle$}
 \put(91,151){$E_{000}$} \put(91,131){$E_{111}$}
 \put(91,111){$E_{011}$} \put(91,91){$E_{100}$}
 \put(91,71){$E_{001}$} \put(91,51){$E_{110}$}
 \put(91,31){$E_{010}$} \put(91,11){$E_{101}$}
 \put(49,133){$E_{00}$} \put(49,93){$E_{11}$}
 \put(49,53){$E_{01}$} \put(49,13){$E_{10}$}
 \put(18,32){$E_{1}$} \put(19,112){$E_{0}$}
 \put(3,78){$1$}
 \newsavebox{\stageIII}
 \savebox{\stageIII}(40, 20)[l]{
   \put(0, 10){\line(4, 1){40}}
   \put(0, 10){\line(4, -1){40}}
   \put(0, 10){\circle*{2}}
   \put(40, 0){\circle*{2}}
   \put(40, 20){\circle*{2}}
 }
 \newsavebox{\stageII}
 \savebox{\stageII}(40, 40)[l]{
   \put(0, 20){\line(3, 2){30}}
   \put(0, 20){\line(3, -2){30}}
   \put(0, 20){\circle*{2}}
 }
 \newsavebox{\stageI}
 \savebox{\stageI}(40, 40)[l]{
   \put(0, 40){\line(1, 2){20}}
   \put(0, 40){\line(1, -2){20}}
   \put(0, 40){\circle*{2}}
 }
 \multiput(50,10)(0,40){4}{\usebox{\stageIII}}
 \multiput(20,20)(0,80){2}{\usebox{\stageII}}
 \multiput(0,60)(0,80){1}{\usebox{\stageI}}
 \end{picture}
 \end{center}
 \caption{Labeling tree the Fourier basis elements. The nodes in level $k$ are labeled
 with the $2^k\times 2^k$ representations $E_{\tau_k (g)}$ of $u\in H_k/\langle d_k \rangle$.
\label{LabelBasis}}
\end{figure}
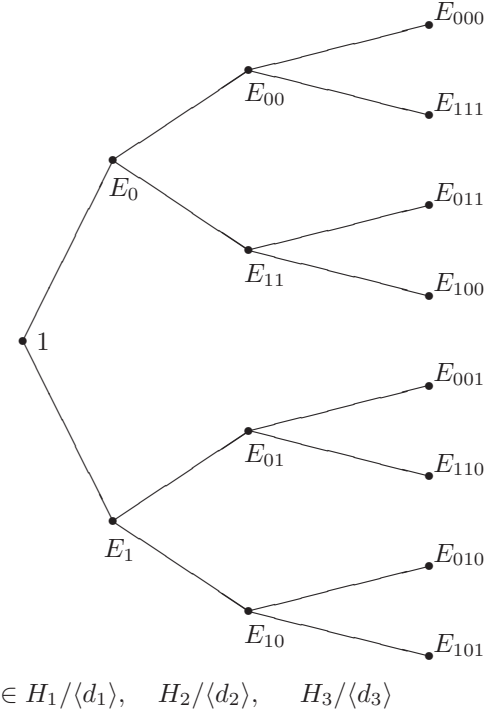
 \begin{figure}
 \begin{center}
 \setlength{\unitlength}{0.6mm}
 \begin{picture}(120, 160)(-9,0)
 \put(-6,0){$v\in G, ~~{G}/{H_1}, ~~~{G}/{H_2}  $}
 \put(91,151){${000},f_0$} \put(91,131){${100},f_4$}
 \put(91,111){${010},f_2$} \put(91,91){${110},f_6$}
 \put(91,71){${001},f_1$} \put(91,51){${101},f_5$}
 \put(91,31){${011},f_3$} \put(91,11){${111},f_7$}

 \put(48,132){${b_1 00}$} \put(48,92){${b_1 10}$}
 \put(48,52){${b_1 01}$} \put(48,12){${b_1 11}$}

 \put(11,32){${b_1b_21}$} \put(11,125){${b_1b_20}$}
 \put(6,84){$b_1 b_2 b_3$}

 \put(45,147){$f_0+f_4$} \put(45,106){$f_2+f_6$}

 \put(45,66){$f_1+f_5$} \put(45,27){$f_3+f_7$}

 \put(-15,23){$f_1+f_3+f_5+f_7$} \put(-15,135){$f_0+f_2+f_4+f_6$}
 \put(8,72){$\sum_0^7\! f_j$}

 \savebox{\stageIII}(40, 20)[l]{
   \put(0, 10){\line(4, 1){40}}
   \put(0, 10){\line(4, -1){40}}
   \put(0, 10){\circle*{2}}
   \put(40, 0){\circle*{2}}
   \put(40, 20){\circle*{2}}
 }

 \savebox{\stageII}(40, 40)[l]{
   \put(0, 20){\line(3, 2){30}}
   \put(0, 20){\line(3, -2){30}}
   \put(0, 20){\circle*{2}}
 }

 \savebox{\stageI}(40, 40)[l]{
   \put(0, 40){\line(1, 2){20}}
   \put(0, 40){\line(1, -2){20}}
   \put(0, 40){\circle*{2}}
 }
 \multiput(50,10)(0,40){4}{\usebox{\stageIII}}
 \multiput(20,20)(0,80){2}{\usebox{\stageII}}
 \multiput(0,60)(0,80){1}{\usebox{\stageI}}
 \end{picture}
 \end{center}
 \caption{Labeling tree of the arguments of $f$ in the sums in (\ref{eq_FT}) that multiply the
 Fourier basis elements. The bit vector labels of the elements in $G$ are obtained by letting
 $b_1,b_2$, and  $b_3$ vary in $\{0,1\}$. \label{LabelCoeffs}}
 \end{figure}
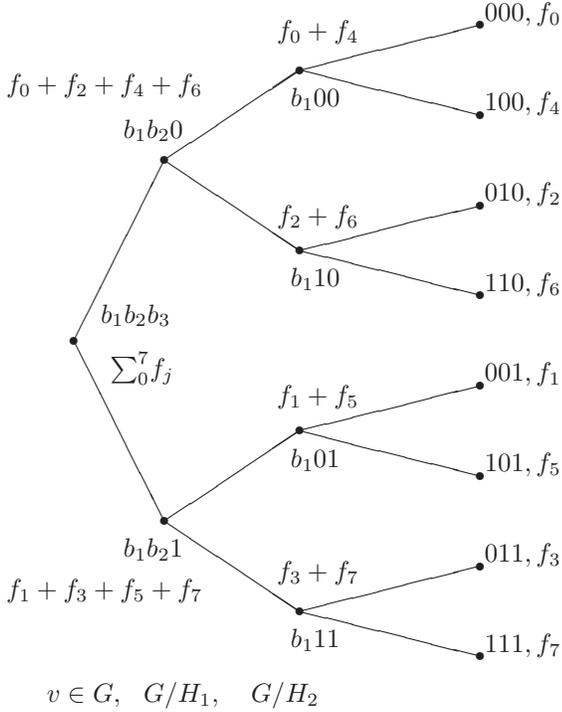
 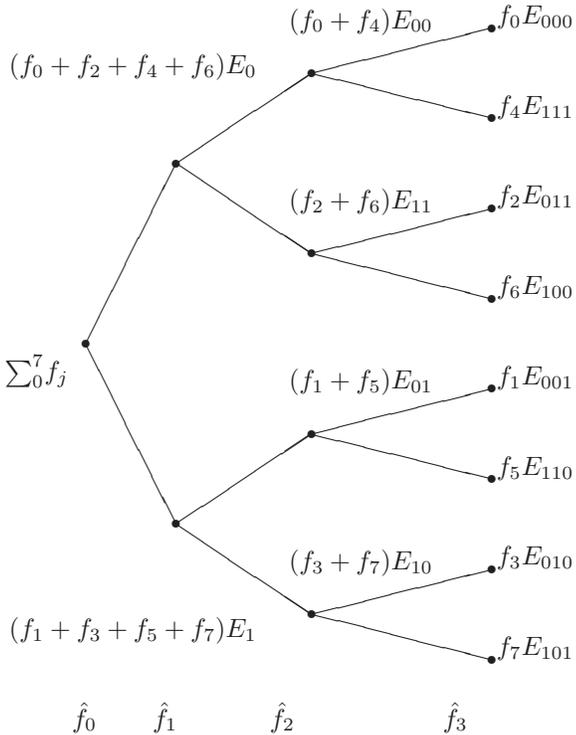
\begin{figure}
 \begin{center}
 \setlength{\unitlength}{0.6mm}
 \begin{picture}(120, 160)(-7,-5)
 \put(-3,-5){$\hat{f}_0 ~~~~~~\hat{f}_1 ~~~~~~~~~~\hat{f}_2   ~~~~~~~~~~~~~~~~\hat{f}_3$}
 \put(91,151){$f_0 E_{000}$} \put(91,131){$f_4 E_{111}$}
 \put(91,111){$f_2 E_{011}$} \put(91,91){$f_6 E_{100}$}
 \put(91,71){$f_1 E_{001}$} \put(91,51){$f_5 E_{110}$}
 \put(91,31){$f_3 E_{010}$} \put(91,11){$f_7 E_{101}$}



 \put(45,150){$(f_0+f_4)E_{00}$} \put(45,110){$(f_2+f_6)E_{11}$}

 \put(45,70){$(f_1+f_5)E_{01}$} \put(45,30){$(f_3+f_7)E_{10}$}

 \put(-17,15){$(f_1+f_3+f_5+f_7)E_1$} \put(-17,140){$(f_0+f_2+f_4+f_6)E_0$}
 \put(-18,72){$\sum_0^7\! f_j$}

 \savebox{\stageIII}(40, 20)[l]{
   \put(0, 10){\line(4, 1){40}}
   \put(0, 10){\line(4, -1){40}}
   \put(0, 10){\circle*{2}}
   \put(40, 0){\circle*{2}}
   \put(40, 20){\circle*{2}}
 }

 \savebox{\stageII}(40, 40)[l]{
   \put(0, 20){\line(3, 2){30}}
   \put(0, 20){\line(3, -2){30}}
   \put(0, 20){\circle*{2}}
 }

 \savebox{\stageI}(40, 40)[l]{
   \put(0, 40){\line(1, 2){20}}
   \put(0, 40){\line(1, -2){20}}
   \put(0, 40){\circle*{2}}
 }
 \multiput(50,10)(0,40){4}{\usebox{\stageIII}}
 \multiput(20,20)(0,80){2}{\usebox{\stageII}}
 \multiput(0,60)(0,80){1}{\usebox{\stageI}}
 \end{picture}
 \end{center}
 \caption{Labeling tree of the multiplications of the arguments of $f$ in the sums in (\ref{eq_FT}) and the corresponding
 $E$ matrices, resulting in the fast TMFT coefficients. \label{LabelFastTMFT}}
 \end{figure}

In Fig.~\ref{LabelBasis}, we illustrate how to efficiently bit-label all the $E$ matrices
using a binary tree structure. At level $k$ in the tree, let {\color{black} ${{\bf b}^{(k)}}=\tau_k (g)$} denote a $k$ bit vector, then
\[
E_{{\bf b}^{(k)}}\in \bigcup_{u\in H_k/\langle d_k \rangle} \{E_{\sigma_k(u)}, E_{\overline{\sigma_k(u)}}\} ~~k=1,\ldots, n
\]
{\color{black} where $g=u+v$ or $g=u+v+d_k$ for some $u\in H_k/\langle d_k\rangle$ and $v\in G/{H_k}$}.
At level $k$, a node labeled with $E_{{\bf b}^{(k)}}$ splits into two branches leading to an upper node labeled by $E_{0,{{\bf b}^{(k)}}}$
(prepend 0) and a lower node labeled by $E_{1,\overline{{\bf b}^{(k)}}}$ (complement bits of
${{\bf b}^{(k)}}$ and prepend 1). This pair of nodes with a common parent correspond to
$E_{\sigma_{k+1}(u)}$ and $E_{\overline{\sigma_{k+1}(u)}}$, respectively.

From Definition~\ref{fastTMFT}, we note that the Fourier coefficient matrices $\hat{f}_k$
in the fast TMFT are linear combination of the $E_{{\bf b}_k}$ matrices, weighted by a sum of the time
domain samples of the function $f$ given in (\ref{eq_FT}). Fig.~\ref{LabelCoeffs} illustrates how we label the nodes
for $v\in G/H_{k}$.
For convenience of notation, the time domain samples in Fig.~\ref{LabelCoeffs} are denoted
by $f_j=f(g)$, $j=D(g)$, for any $g\in G$. At level $k$, in each pair of the nodes with a common parent,
the upper binary vector represents the $u+v$  and  the lower represents $u+v+d_k$, where
$v\in G/H_k$ for a given $u$. This tree can be used to compute the sums over $v\in G/H_k$ in (\ref{eq_FT}).

Combining the labels from both trees in Figs.~\ref{LabelBasis} and \ref{LabelCoeffs} yields
the combined tree structure illustrated in Fig.~\ref{LabelFastTMFT}. For each pair of nodes
at level $k$ with the same parent node at level $k-1$, we compute
\begin{equation}\label{partI_FT}
\left[\sum_{v\in G/H_k} f(u+v)\right] E_{\sigma_k(u)}
\end{equation}
and
\begin{equation}\label{partII_FT}
\left[\sum_{v\in G/H_k}f(u+d_k+v) \right] E_{\overline{\sigma_k(u)}}
\end{equation}
respectively,
where $E_{\sigma_k(u)}$ and $E_{\overline{\sigma_k(u)}}$ are the node labels from Fig. \ref{LabelBasis},
while the arguments of $f$ in the sums are given by the node labels from Fig. \ref{LabelCoeffs}.
Using Fig.~\ref{LabelFastTMFT}, we can explicitly rewrite the fast TMFT coefficients in (\ref{Ex2FT0}),
(\ref{Ex2FT1}), (\ref{Ex2FT2}), and (\ref{Ex2FT3}) as
\begin{eqnarray}\label{exactFC}
\hat{f}_0 &=&  f_0 + f_1 + f_2 + f_3 + f_4 + f_5 + f_6 + f_7 \nonumber\\
\hat{f}_1 &=&  (f_0 + f_2 + f_4 + f_6) E_0 + (f_1 + f_3 + f_5 + f_7) E_1 \nonumber\\
\hat{f}_2 &=&  (f_0 + f_4 ) E_{00} + (f_2 + f_6 ) E_{11} \nonumber\\
&& +(f_1 + f_5) E_{01} + (f_3 + f_7) E_{10} \nonumber\\
\hat{f}_3 &=&  f_0 E_{000} + f_4 E_{111} + f_2 E_{011} + f_6 E_{100} \nonumber\\
&& +f_1 E_{001} + f_5 E_{110} + f_3E_{010} + f_7 E_{101}
\end{eqnarray}

Alternatively, the Fourier basis vectors ${\boldsymbol\psi}_k$ in Definition~\ref{Def:absFT} (non-fast TMFT) are shown in
Table \ref{ex:basisvectors}.
\begin{table*}[t]
\begin{center}\caption{The Fourier basis vectors for $G=C_2^3$ \label{ex:basisvectors}}
\begin{tabular} {|c||cccccccc||c|}
\hline
$D(g)$           & $0$ & $1$ & $2$ & $3$ & $4$ & $5$ &  $6$ & $7$ &~ \\ \hline\hline
$E_{\tau_0(g)}$ & $1$ & $1$ & $1$ & $1$ & $1$ & $1$ & $1$ & $1$ & ${\boldsymbol\psi}_0$ \\
$E_{\tau_1(g)}$ & $E_0$ & $E_1$ & $E_0$ & $E_1$ & $E_0$ & $E_1$ & $E_0$ & $E_1$ &
  ${\boldsymbol\psi}_1$ \\
$E_{\tau_2(g)}$ & $E_{00}$ & $E_{01}$ & $E_{11}$ & $E_{10}$ & $E_{00}$ & $E_{01}$ & $E_{11}$ & $E_{10}$  & ${\boldsymbol\psi}_2$ \\
$E_{\tau_3(g)}$ & $E_{000}$ & $E_{001}$ & $E_{011}$ & $E_{010}$ & $E_{111}$ & $E_{110}$ & $E_{100}$ & $E_{101}$   & ${\boldsymbol\psi}_3$ \\
\hline
\end{tabular}
\end{center}
\end{table*}
$\hfill\square$
\end{myexample}
\begin{myexample}
The TMFT of a Dirac function over $G$, i.e., $\delta_{0} (0) = 1$ and $0$ otherwise, is given by
\[
\widehat{\delta_{0}(g)} = \left[1, E_0, \ldots, E_{{\bf 0}_n} \right] = \left[1, {\bf I}_2, \ldots, {\bf I}_{2^n} \right]
\]
i.e., the list of Fourier coefficient matrices is made up of identity matrices of increasing size.
 $\hfill\square$
\end{myexample}

\begin{myexample}
The TMFT of the indicator function of the element $g_0 \in G$, i.e.,
\[
\delta_{g_0} (g) = \left\{ \begin{array}{cc}
                     1 & g=g_0 \\
                     0 & \mbox{otherwise}
                   \end{array}\right.
\]
is given by
\[
\widehat{\delta_{g_0} (g)} = \left[1, E_{\tau_1 (g_0)}, \ldots, E_{\tau_n (g_0)} \right].
\]
For example, $g_0 = (11) \in G=C_2^2$ yields $\widehat{\delta_{(11)} (g)}$ = $\left[1, E_{\tau_1 (11)}, E_{\tau_2 (11)} \right]$ = $\left[1, E_1, E_{10}\right]$.  $\hfill\square$
\end{myexample}
%
\section{The inverse two-modular Fourier transform}\label{secITMFT}
%
In the case of binary functions considered in this paper, Definition \ref{def_groupIDFT} cannot
be applied since the $\frac{1}{|G|}$Tr$(\cdot)$ operator is undefined. In fact $|G|$ does not
have an inverse and the trace of any representation matrix $E_{\bf b}$  is always zero, having an even number
of ones on the diagonal (see Property \ref{prop:p5}). To overcome this problem, we introduce the matrix operator $\Phi_k: \pi_k(C_2^k) \rightarrow C_2$
from the set of two-modular representations of $C_2^k$ to $\{0,1\}$, for $k=1,\ldots,n$.
\begin{mydefinition} \label{Def_Phi}
Let $E_{\bf b}$ be the $2^k\times 2^k$ representation of a $k$-bit binary vector
${\bf b}\in C_2^k$, then we define the matrix operator on $E_{\bf b}$ as
\[
\Phi_k(E_{\bf b}) \triangleq  E_{\bf b}[1,2^k]   \in \{0,1\}
\]
i.e.,  $\Phi_k$  extracts the top-right corner element of the matrix $E_{\bf b}$.

$\hfill\square$

\end{mydefinition}
As observed in Property \ref{prop:p4}, only ${\bf b}={\bf 1}_k$ yields {\color{black} $\Phi_k(E_{\bf b}) = 1$}, while any other binary vector representation is mapped to zero.
\begin{mylemma}\label{lemma:IFFT_LinearOperator}
The operator $\Phi_k$ is linear, i.e.,
\[
\Phi_k(\alpha E_{\bf a} + \beta E_{\bf b}) = \alpha \Phi_k(E_{\bf a}) + \beta \Phi_k(E_{\bf b})
\]
and
\begin{equation}\label{communitative}
 \Phi_k(E_{\bf a}  E_{\bf b})=\Phi_k(E_{\bf b}  E_{\bf a})
 \end{equation}
for any $\alpha,\beta \in {\cal R}$, and ${\bf a},{\bf b}\in C_2^k$.
\end{mylemma}
{\em Proof}: The proof is straightforward.
\begin{mylemma}\label{lemma:IFFT_E}
Let $E_{\bf a}$ and $E_{\bf b}$ be the $2^k\times 2^k$ representation of the $k$-bit binary vectors ${\bf a},{\bf b}\in C_2^k$, respectively. We have

  \begin{eqnarray}\label{Phi_k}
\Phi_k(E_{\bf a}  E_{\bf b}) &=& \Phi_k(E_{{\bf a} + {\bf b}}) \nonumber\\
&=&
\left\{ \begin{array}{cc}
         1 & \mbox{iff~} {\bf a} + {\bf b} = {\bf 1} ~(\mbox{or~} {\bf a} = \bar{\bf b}) \\
         0 & \mbox{otherwise}~.
 \end{array} \right.
  \end{eqnarray}
 $\hfill\square$
\end{mylemma}
{\em Proof}:
The proof is straightforward.
$\hfill\square$

\begin{mytheorem} ({\bf Inverse TMFT}). The {\em inverse TMFT} is given by
\begin{equation} \label{eq_InvFT}
f_j = \hat{f}_0 + \sum^{n}_{k=1}\Phi_{k}\left(\hat{f}_k {E}_{\tau_k (D^{-1}(j))}   \right)~~
j=0,\ldots, 2^n-1
\end{equation}
where $D^{-1}(j) = (c_n, \ldots, c_k, \ldots, c_1)$ {\footnote{To simplify notation, we have reversed the order of the bit indices of $c_k$.}},
with $c_k \in \{0,1\}$ and $\tau_k$ is given in (\ref{eq:tau_k_u}).
$\hfill\square$
\end{mytheorem}
{\em Proof}: Recalling (\ref{eq:gHk}), (\ref{eq:tree_decomp}), (\ref{eq:Hkdk}), and $d_k=(0,\ldots, b_{n-k+1}=1, 0,\ldots, 0)$ (the $n$-bit all-zero vector except for $b_{n-k+1}=1$),
we have
\begin{eqnarray}  \label{eq:ugdk1}
\left\{ \begin{array}{cl}
u+v & =(b_1, \ldots, b_{n-k}, 0, b_{n-k+2}, \ldots, b_n)   \\
u+v+d_k & =(b_1, \ldots, b_{n-k}, 1, b_{n-k+2}, \ldots, b_n)
\end{array} \right.
\end{eqnarray}
for $k=2,\ldots,n-1$,
while in the special cases of $k=1$ and $k=n$, we have respectively
\begin{eqnarray}  \label{eq:ugdk2}
\begin{array}{c}
  \left\{ \begin{array}{cl}
u+v & =~(b_1, \ldots, b_{n-k}, 0)  \\
u+v+d_1 & =~(b_1, \ldots, b_{n-k}, 1)
\end{array} \right.
\end{array}
\end{eqnarray}
and
\begin{eqnarray}
\begin{array}{c}
\left\{ \begin{array}{cl}
u+v & =~(0,b_2, \ldots, b_n) \\
u+v+d_n & =~(1,b_2, \ldots, b_n)
\end{array} \right.
\end{array}
\end{eqnarray}
We then rewrite the right-hand side of (\ref{eq_InvFT}) in its binary form as (\ref{eq:nonfast_ITMFT_binary}).
\newcounter{tempequationcounter}
\begin{figure*}[!t]
\normalsize
\begin{IEEEeqnarray}{rCl}
\setcounter{equation}{40}
\lefteqn{\hat{f}_0 + \sum^{n}_{k=1}\Phi_{k}\left(\hat{f}_k {E}_{\tau_k (D^{-1}(j))}\right)}\nonumber\\
&=& \sum_{b_1,\ldots,b_n}^{(v)} f(b_1,\ldots, b_n) + \sum_{b_1,\ldots,b_{n-1}}^{(v)}  \left\{ f(b_1,\ldots, b_{n-1},\underbrace{0}_{c_1}) \Phi_{1}(E_{\sigma_1 (u) =0}{E}_{\tau_1 (D^{-1}(j))}) \right.\nonumber\\
&&\hspace{4.9cm}\left.+ f(b_1,\ldots, b_{n-1},\underbrace{1}_{c_1}) \Phi_{1}(E_{\overline{\sigma_1(u)} = 1}{E}_{\tau_1 (D^{-1}(j))}) \right\}+\cdots\nonumber\\
&+& \sum_{b_1,\ldots,b_{n-k}}^{(v)} \sum_{b_{n-k+2},\ldots,b_n}^{(u)}  \left\{f(b_1,\ldots, b_{n-k},\underbrace{0}_{c_k}, b_{n-k+2},\ldots, b_{n}) \Phi_{k}(E_{\sigma_k(u)}{E}_{\tau_k (D^{-1}(j))}) \right.\nonumber\\
&&\hspace{3.2cm}\left.+ f(b_1,\ldots, b_{n-k},\underbrace{1}_{c_k}, b_{n-k+2},\ldots, b_{n}) \Phi_{k}(E_{\overline{\sigma_k(u)}}{E}_{\tau_k (D^{-1}(j))}) \right\}+\cdots\nonumber\\
&+& \sum_{b_2,\ldots,b_n}^{(u)} \left\{f(\underbrace{0}_{c_n},b_2,\ldots, b_{n}) \Phi_{n}(E_{\sigma_n(u)}{E}_{\tau_n (D^{-1}(j))}) + f(\underbrace{1}_{c_n},b_2,\ldots, b_{n}) \Phi_{n}(E_{\overline{\sigma_n(u)}}{E}_{\tau_n (D^{-1}(j))})\right\}.\nonumber\\
\label{eq:nonfast_ITMFT_binary}
\end{IEEEeqnarray}
\setcounter{equation}{40}
\hrulefill
\vspace*{4pt}
\end{figure*}

For any $k\in\{1,\ldots,n\}$, based on the binary representation $D^{-1}(j) = (c_n, \ldots, c_k, \ldots, c_1)$ and
the definition of $\tau_k$ in (\ref{eq:tau_k_u}), we have that:
\begin{itemize}
\item
${E}_{\tau_k (D^{-1}(j))} = E_{\sigma_k ({\tilde u})}$ holds when $D^{-1}(j) \in \{{\tilde u}+v~| ~v\in G/H_k\}$,
which implies $c_k=0$ and  ${\tilde u}=(0,\ldots,0,c_{k-1},\ldots, c_1)$.
For such $\tilde{u}$, only the term $f(b_1,\ldots, b_{n-k},\bar{c}_k=1, c_{k-1},\ldots, c_1)$ in the sum over $u$ remains,
since $\Phi_{k}(E_{\overline{\sigma_k({\tilde u})}}{E}_{\tau_k (D^{-1}(j))}) = \Phi_{k}(E_{\overline{\sigma_k({\tilde u})}}E_{\sigma_k ({\tilde u})})=1$.
All the other terms cancel, since
$\Phi_{k}(E_{\sigma_k(u)}E_{\sigma_k ({\tilde u})})=0$ for all $u$, and
$\Phi_{k}(E_{\overline{\sigma_k(u)}}E_{\sigma_k ({\tilde u})})=0$ for all $u\neq \tilde{u}$.

\item
${E}_{\tau_k (D^{-1}(j))} = E_{\overline{\sigma_k ({\tilde u})}}$ holds when $D^{-1}(j) \in \{{\tilde u}+v+d_k~| ~v\in G/H_k\}$, which implies $c_k=1$ and ${\tilde u}=(0,\ldots,0,c_{k-1},\ldots, c_1)$.
For such $\tilde{u}$, only the term $f(b_1,\ldots, b_{n-k},\bar{c}_k=0, c_{k-1},\ldots, c_1)$ in the sum over $u$ remains,
since $\Phi_{k}(E_{\sigma_k(\tilde{u})}E_{\overline{\sigma_k({\tilde u})}})=1$, while the other terms cancel,
since $\Phi_{k}(E_{\overline{\sigma_n(u)}}E_{\overline{\sigma_k({\tilde u})}})=0$
for all $u$ and $\Phi_{k}(E_{\sigma_k(u)}E_{\overline{\sigma_k({\tilde u})}})=0$ for all $u\neq \tilde{u}$.
\end{itemize}

Then (\ref{eq:nonfast_ITMFT_binary}) simplifies to
\begin{eqnarray}
&&\sum_{b_1,\ldots,b_n} f(b_1,\ldots, b_n) \nonumber\\
&+& \sum_{b_1,\ldots,b_{n-1}}^{(v)}  f(b_1,\ldots, b_{n-1},\bar{c}_1) + \cdots \label{eq:simplify_ITMFT_binary1}\\
&+& \sum_{b_1,\ldots,b_{n-k}}^{(v)} f(b_1,\ldots, b_{n-k},\bar{c}_k, c_{k-1},\ldots, c_{1})  \label{eq:simplify_ITMFT_binary3}\\
&+& \cdots + f(\bar{c}_n, {c}_{n-1},\ldots, c_{1})~. \label{eq:simplify_ITMFT_binary4}
\end{eqnarray}

Adding the first two summations in (\ref{eq:simplify_ITMFT_binary1}) yields
$\sum_{b_1,\ldots,b_{n-1}} f(b_1,\ldots, b_{n-1},{c}_1)$
due to the characteristic 2 of ${\cal R}$ (bitwise XOR addition).
Progressively adding the summations up to (\ref{eq:simplify_ITMFT_binary3})
yields $\sum_{b_1,\ldots,b_{n-k}} f(b_1,\ldots, b_{n-k},c_k,\ldots, c_1)$.
Finally adding all summations up to (\ref{eq:simplify_ITMFT_binary4}) yields
$f(c_n,\ldots, c_k,\ldots, c_1) = f_j$.
This completes the proof.

$\hfill\square$

\begin{myexample}
Following Example \ref{exampleFFT}, given Fourier coefficients
$\hat{f}_0$, $\hat{f}_1$, $\hat{f}_2$ and $\hat{f}_3$ in (\ref{exactFC}) for a function over $G=C_2^3$, the
inverse Fourier transform can be computed as
\begin{eqnarray*}
{f}_0 \hspace{-3mm}&=&\hspace{-4mm} \hat{f}_0 + \Phi_1(\hat{f}_1 E_0) + \Phi_2(\hat{f}_2 E_{00}) + \Phi_3(\hat{f}_3 E_{000})\\
{f}_1 \hspace{-3mm}&=&\hspace{-4mm} \hat{f}_0 + \Phi_1(\hat{f}_1 E_1) + \Phi_2(\hat{f}_2 E_{01}) + \Phi_3(\hat{f}_3 E_{001})\\
{f}_2 \hspace{-3mm}&=&\hspace{-4mm} \hat{f}_0 + \Phi_1(\hat{f}_1 E_0) + \Phi_2(\hat{f}_2 E_{11}) + \Phi_3(\hat{f}_3 E_{011})\\
{f}_3 \hspace{-3mm}&=&\hspace{-4mm} \hat{f}_0 + \Phi_1(\hat{f}_1 E_1) + \Phi_2(\hat{f}_2 E_{10}) + \Phi_3(\hat{f}_3 E_{010})\\
{f}_4 \hspace{-3mm}&=&\hspace{-4mm} \hat{f}_0 + \Phi_1(\hat{f}_1 E_0) + \Phi_2(\hat{f}_2 E_{00}) + \Phi_3(\hat{f}_3 E_{111})\\
{f}_5 \hspace{-3mm}&=&\hspace{-4mm} \hat{f}_0 + \Phi_1(\hat{f}_1 E_1) + \Phi_2(\hat{f}_2 E_{01}) + \Phi_3(\hat{f}_3 E_{110})\\
{f}_6 \hspace{-3mm}&=&\hspace{-4mm} \hat{f}_0 + \Phi_1(\hat{f}_1 E_0) + \Phi_2(\hat{f}_2 E_{11}) + \Phi_3(\hat{f}_3 E_{100})\\
{f}_7 \hspace{-3mm}&=&\hspace{-4mm} \hat{f}_0 + \Phi_1(\hat{f}_1 E_1) + \Phi_2(\hat{f}_2 E_{10}) + \Phi_3(\hat{f}_3 E_{101})
\end{eqnarray*}
$\hfill\square$
\end{myexample}
%
\section{TMFT Properties}\label{secConv}
%
\begin{mytheorem}
({\bf Linearity of TMFT}).~Given a pair of functions $r$ and $s: G \rightarrow {\cal R}$, let
$\hat{r}=[\hat{r}_0, \ldots, \hat{r}_k, \ldots, \hat{r}_n]$ and
$\hat{s}=[\hat{s}_0, \ldots, \hat{s}_k, \ldots, \hat{s}_n]$
be the lists of Fourier coefficients matrices of TMFT (i.e., $\hat{r}_k$ and $\hat{s}_k$
are $2^k\times 2^k$ matrices), the TMFT of the linear combination of $r$ and $s$ is given by
\begin{eqnarray*}
\widehat{\alpha r + \beta s } &=& \alpha \hat{r} + \beta \hat{s} \nonumber\\
&=& [\alpha\hat{r}_0+\beta\hat{s}_0, \ldots, \alpha\hat{r}_k+\beta\hat{s}_k, \ldots, \alpha\hat{r}_n+\beta\hat{s}_n]
\end{eqnarray*}
for $\alpha,\beta\in{\cal R}$.
 $\hfill\square$
\end{mytheorem}

{\em Proof}: The proof is straightforward.

Next, we specialize the definition of convolution in (\ref{EQconv}) for the case of the
additive group $G=C_2^n$.
\begin{mydefinition}\label{def:conaddgroup}
Given a pair of functions $r$ and $s: G \rightarrow {\cal R}$ we define the
convolution product $f: G \rightarrow {\cal R}$ as
\[
f(g) =  r(g) \ast s(g) = \sum_{g'\in G} r(g'+g) s(g')~~ \mbox{for} ~g\in G.
\]
~~~~~~~~~~~$\hfill\square$
\end{mydefinition}
It can be easily shown that the convolution product is commutative.
Now we present the convolution theorem when using TMFT.
\begin{mytheorem}
({\bf Convolution Theorem}).~Given a pair of functions $r$ and $s: G \rightarrow {\cal R}$, let
$\hat{r}=[\hat{r}_0, \ldots, \hat{r}_k, \ldots, \hat{r}_n]$ and
$\hat{s}=[\hat{s}_0, \ldots, \hat{s}_k, \ldots, \hat{s}_n]$
be the lists of Fourier coefficients matrices of TMFT (i.e., $\hat{r}_k$ and $\hat{s}_k$
are $2^k\times 2^k$ matrices), we obtain Fourier transform of the convolution product
as
\[
\widehat{r \ast s } = \hat{r} \odot \hat{s} \triangleq [\hat{r}_0\hat{s}_0, \ldots, \hat{r}_k\hat{s}_k, \ldots, \hat{r}_n\hat{s}_n]~.
\]
~~$\hfill\square$
\end{mytheorem}

{\em Proof}:
Using Definition \ref{Def:absFT}, we can simply write the product of the $k$-th Fourier coefficient matrices of $r$ and $s$ as
\[
\hat{r}_k \hat{s}_k = \sum_{g\in G} r(g) E_{\tau_k(g)}   \sum_{g'\in G} s(g') E_{\tau_k(g')}~.
\]
Substituting $w = g + g'$, we obtain
\begin{eqnarray} \label{eq:rs1}
\hat{r}_k \hat{s}_k &=& \sum_{w\in G} \sum_{g\in G} r(g) s(g+w) E_{\tau_k(g)}E_{\tau_k(g+w)} \nonumber\\
&=& \sum_{w\in G} \sum_{g\in G} r(g) s(g+w) E_{\tau_k(g)}E_{\tau_k(g)+\tau_k(w)}\nonumber\\
&=& \sum_{w\in G} \sum_{g\in G} r(g) s(g+w) E_{\tau_k(w)} \nonumber\\
&=& \sum_{w\in G} (r\ast s)(w)E_{\tau_k(w)}~.
\end{eqnarray}
The second equality is based on the fact that $\tau_k$ is group homomorphism (see Lemma~\ref{lemma:tau_kgrouphomo}).
~~~$\hfill\square$

\begin{mytheorem}
({\bf Shifting Property}).~Given the function $f: G \rightarrow {\cal R}$ and its TMFT
\[
\hat{f}_k = \sum_{g\in G} f(g) E_{\tau_k (g)} ~~k=1,\ldots, n
\]
and a given shift $a\in G$ then the Fourier transform of $f(g+a)$ is given by
\begin{equation}\label{timeshift1}
\sum_{g\in G} f(g+a) E_{\tau_k (g+a)} = \sum_{g\in G} f(g+a) E_{\tau_k (g)}E_{\tau_k (a)}~.
\end{equation}
If $f(g+a)=f(g)$ for all $g\in G$, then the above Fourier transform becomes
\begin{equation}\label{timeshift2}
\sum_{g\in G} f(g+a) E_{\tau_k (g+a)} = \hat{f}_k E_{\tau_k (a)}~.
\end{equation}
~~$\hfill\square$
\end{mytheorem}

{\em Proof}: The proof is straightforward.
%
\section{Implementation and Complexity}\label{complexity}
%
The evaluation of the TMFT only requires additions (and no multiplications) in the
ring ${\cal R}$, since the $E_{\bf b}$ matrices only contain zeros and ones in ${\cal R}$.
Hence, we define the {\em complexity} as the number of additions in the ring ${\cal R}$.
For convenience of exposition, we will begin by analyzing the complexity of the ITMFT.

\subsection{Complexity of ITMFT}

The following lemma enables us to count the number of ring additions needed to compute
one term $\Phi_k(\hat{f}_k E_{\tau_{k} (D^{-1}(j))})$ in (\ref{eq_InvFT}), for any
$k=0,\ldots, n$ and $j=0,\ldots, 2^n-1$.
We note that the top right corner of the matrix product is given by the scalar product
first row of $\hat{f}_k$ and the last column of $E_{\tau_{k} (D^{-1}(j))}$.

\begin{mylemma} \label{KronHammingWeight}
Given an $n$-bit vector ${\bf b}$ and the corresponding representation matrix $E_{\bf b}$,
let ${\bf v}$ be the first row (or the transposed last column) of the matrix $E_{\bf b}$ and let $w_H({\bf b})$ and $w_H({\bf v})$
be their Hamming weights, then
\begin{equation} \label{eq:weight_lastcol}
w_H({\bf v}) = 2^{w_H({\bf b})}~.
\end{equation}

$\hfill\square$
\end{mylemma}

{\em Proof:}
We first prove this lemma when ${\bf v}$ is the first row of the matrix $E_{\bf b}$.
For $n=1$, (\ref{eq:weight_lastcol}) is true by definition of $E_0$ and $E_1$.
By induction on the number of bits, we assume (\ref{eq:weight_lastcol})
is true for a $k$-bit vector
${\bf b}^{(k)}$, i.e., $w_H({\bf v}^{(k)}) = 2^{w_H({\bf b}^{(k)})}$,
where ${\bf v}^{(k)}$ is the first row of $E_{{\bf b}^{(k)}}$.
When one more bit $b_{k+1}$ is appended to ${\bf b}^{(k)}$ the matrix
representation becomes
\[
E_{{\bf b}^{(k+1)}} = E_{{\bf b}^{(k)}} \otimes E_{b_{k+1}}~.
\]
From the definition of the Kroneker product and the matrices $E_0$ and $E_1$,
we have:
\[
w_H({\bf v}^{(k+1)}) =
\left\{\begin{array}{cc}
  w_H({\bf v}^{(k)}) & \mbox{if}~b_{k+1}=0\\
2 w_H({\bf v}^{(k)}) & \mbox{if}~b_{k+1}=1~
\end{array} \right.
\]
Hence the weight of the first row doubles for every bit that is equal to one in ${\bf b}$.

Based on the anti-diagonal symmetry noted in Property \ref{prop:p1},
under the same assumptions, (\ref{eq:weight_lastcol})
is also valid when ${\bf v}^T$ is the last column of $E_{\bf b}$.
$\hfill\square$

\begin{mylemma}
The total complexity of the ITMFT is given by
\begin{eqnarray}
C_{\mbox{ITMFT}}  = \frac{3^{n+1} +1}{2} + (n-2)2^n~. \label{ComplexityITMFT2}
\end{eqnarray}$\hfill\square$
\end{mylemma}

{\em Proof}:
The total complexity of the ITMFT takes into accounts {\em i)} the number of terms in $\hat{f}_k$ to be added when
computing $\Phi_k(\hat{f}_k E_{\tau_{k} (D^{-1}(j))})$, for $k=1, \ldots, n$;
and {\em ii)} the number of additions of terms $\Phi_k(\hat{f}_k E_{\tau_{k} (D^{-1}(j))}) $ in (\ref{eq_InvFT}).

Let $w=w_H({\bf b})$ be the Hamming weight of the $k$-bit vector ${\bf b}$ associated with the matrix $E_{\tau_{k} (D^{-1}(j))}$.
The number of elements of the matrix $\hat{f}_k$ to be added when computing $\Phi_k(\hat{f}_k E_{\tau_{k} (D^{-1}(j))})$
is determined by the number of ones in the last column of $E_{\tau_{k} (D^{-1}(j))}$,
which is $2^w$ according to Lemma \ref{KronHammingWeight}. Then the number of additions is one less, i.e., $2^w-1$.
Since there are only $2^k$ distinct $E_{\tau_{k} (D^{-1}(j))}$ for each $k$,
we need to run over all the weights $w$ of the $k$-bit vector corresponding to the matrix $E_{\tau_{k} (D^{-1}(j))}$ for $k=1,\ldots,n$.
This results in a complexity of
\begin{equation}\label{eq:C1_ITMFT}
\sum_{k=1}^n \sum_{w=0}^k {k \choose w} (2^w-1)~.
\end{equation}
We simplify (\ref{eq:C1_ITMFT}) to
\begin{equation}\label{off-diagonal}
\sum_{k=1}^n \left[\sum_{w=0}^k {k \choose w} 2^w-\sum_{w=0}^k  {k \choose w}\right] = \sum_{k=1}^n (3^k -2^k)~.
\end{equation}

On the other hand, the number of additions of terms $\Phi_k(\hat{f}_k E_{\tau_{k} (D^{-1}(j))})$ in (\ref{eq_InvFT}) is $n2^n$.
Finally, we obtain the total complexity
\begin{equation}
C_{\mbox{ITMFT}} = \sum_{k=1}^n (3^k-2^k) + n2^n = \frac{3^{n+1} +1}{2}-2^{n+1} + n2^n~.
\end{equation}
$\hfill\square$

\subsection{Complexity of the fast TMFT}

\begin{figure*}[t]
\begin{center}
\begin{tabular}{ll} \hline
 1.&  {\bf Input}: ${\bf v}$ (first row of $\hat{f}_k$), $k$ number of bits \\
 2.&  {\bf for} $j=0:k-1$ \\
 3.&  \hspace{5mm} ${\bf w} = \mbox{\bf zeros}(2^j,2^k) $;\\
 4.&  \hspace{5mm} {\bf for} $i=1:2^{j+1}:2^k-2^j$\\
 5.&  \hspace{10mm} ${\bf w}(1:2^j,(i+2^j):(i+2^j+2^j-1))=   {\bf v}(1:2^j,i:i+2^j-1)$; \\
 6.&  \hspace{5mm} {\bf end} \\
 7.&  \hspace{5mm}     ${\bf v} = [{\bf v}; {\bf w}]$;\\
 8.&  {\bf end} \\
 9.&  {\bf return} ${\bf v}$ (complete matrix $\hat{f}$)\\ \hline
\end{tabular}
\end{center}
\caption{Algorithm to find the full $\hat{f}_k$ from its first row.
\label{alg_fhat_from_row}}
\end{figure*}

From Lemma \ref{fastTMFT}, we note that the Fourier coefficient matrices $\hat{f}_k$ of the fast TMFT
are linear combination of the matrices $E_{\bf b}$, weighted by the scalar values.
Following (\ref{eq:def_Eb}) and Property \ref{prop:p1}, the matrices $E_{\bf b}$ are the $k$-fold Kroneker products
of the $2\times 2$ upper triangular and anti-diagonal symmetric matrices $E_0$, $E_1$.
This provides a simple algorithm (see Fig. \ref{alg_fhat_from_row}), in which any $\hat{f}_k$ can be entirely reconstructed from its first row entries.
Hence, we only need to compute and store the first row of the matrices $\hat{f}_k$, which is a linear combination of the first rows of matrices $E_{\bf b}$.
Then the complexity can be derived by counting the Hamming weights of the first rows of matrices $E_{\bf b}$.
We have the following Lemma.
\begin{mylemma}
The total complexity of the fast TMFT is given by
\begin{eqnarray}
C_{\mbox{FTMFT}} &=& \frac{3^{n+1} +1}{2}-2^{n+1}~. \label{ComplexityTMFT2}
\end{eqnarray}
$\hfill\square$
\end{mylemma}

{\em Proof}: For each $\hat{f}_k$, $k=1,\ldots, n$, given $k$ bit vector ${\bf b}$ with Hamming weight $w=w_H({\bf b})$, we let ${\bf v_b}$
denote the first row vector of a $E_{\bf b}$ matrix with Hamming weight $w_H({\bf v_b})=2^w$, according to Lemma~\ref{KronHammingWeight}.
We prove the complexity of the fast TMFT in the following steps.
\begin{enumerate}
  \item We start from the leaf nodes at level $n$ in the tree, as shown for example in Fig. \ref{LabelFastTMFT}.
The total number of terms to be added is given by the sum of the Hamming weights of all vectors ${\bf v_b}$ at level $n$, i.e.,
\[
\sum_{{\bf b}} w_H({\bf v_b})=\sum_{w=0}^n  {n \choose w} 2^{w} =  3^n~.
\]
The corresponding addition count is given by
\begin{equation}\label{eq:C1_FTMFT}
\mbox{K}_1 = 3^n -2^n
\end{equation}
since we have $2^n$ separate sums to compute the first row elements.
By direct computation, we note that the first term in the first row of $\hat{f}_n$ is $\hat{f}_0=\sum_{j=0}^{2^n-1} f_j$. This needs to be computed only once and is used throughout the following steps.
\item
At level $k$ ($1\le k <n$) in the tree, we only focus on the first row of each matrix $\hat{f}_k$, except for the first element in this row ($\hat{f}_0=\sum_{j=0}^{2^n-1} f_j$), which has been already computed at level $n$.
Since the first term in ${\bf v_b}$ is always one (see Property \ref{prop:p5}), for each $\hat{f}_k$, the total number of terms
to be added is given by the sum of $w_H({\bf v_b})-1$ of all vectors ${\bf v_b}$ at level $k$, i.e.,
\begin{equation}\label{eq:sumallvector_FTMFT}
\sum_{{\bf b}} (w_H({\bf v_b})-1)=\sum_{w=0}^k  {k \choose w} (2^{w}-1) =  3^k -2^k,
\end{equation}
for $k=1,\ldots,n-1$.
The last equality is due to (\ref{off-diagonal}). The corresponding additions count is given by
$ (3^k -2^k) - (2^k-1)$,
since we have $2^k-1$ separate sums to compute the $2^k-1$ elements of the first row.

At each level $k$, there are extra addition operations that are performed to compute the partial
sum of the time domain samples, i.e., $\sum_{v\in G/H_k} f(u+v)$ in (\ref{partI_FT}) and $\sum_{v\in G/H_k} f(u+v+d_k)$ in (\ref{partII_FT}).
Note that we can ignore the partial sum coefficient of $E_{\bf 0}$ at level $k$, since, by
excluding the first element of the first row of $E_{\bf 0}$,
the remaining elements are all zeros. Hence, the extra count for such addition is $2^k-1$.
This can be also interpreted using the tree structure in Fig. \ref{LabelFastTMFT}: the number of additions
simply coincides with the number of nodes at level $k$, after excluding node $E_{\bf 0}$. Then, the complexity at all level $k$ is given by
\begin{equation}\label{eq:C2_FTMFT}
\mbox{K}_2 =\sum_{k=1}^{n-1} (3^k -2^k) - (2^k-1) + (2^k-1) = \sum_{k=1}^{n-1} (3^k -2^k)~.
\end{equation}
\item
At $k=0$, we have $\hat{f}_0$, already available at level $n$.
Hence, the final complexity is
\[
C_{\mbox{FTMFT}}=\mbox{K}_1 +\mbox{K}_2=\sum_{k=1}^{n} (3^k -2^k)=\frac{3^{n+1} +1}{2}-2^{n+1} ~.
\]
\end{enumerate}
$\hfill\square$

\subsection{Complexity of TMFT}
{\color{black}
\begin{mylemma}
The total complexity of TMFT is given by
\begin{eqnarray}
C_{\mbox{TMFT}} &=& 3^{n+1} - (n+4) 2^n + n + 1~. \label{ComplexitynonfastTMFT}
\end{eqnarray}
$\hfill\square$
\end{mylemma}

{\em Proof}:
The proof is similar to that of the fast TMFT and can be derived by modifying (\ref{eq:sumallvector_FTMFT}) and (\ref{eq:C2_FTMFT}).
\begin{enumerate}
  \item At level $n$, the complexity of TMFT is the same as $\mbox{K}_1=3^n-2^n$ in (\ref{eq:C1_FTMFT}) of the fast TMFT, since
both methods have the same $2^n$ distinct matrices $E_{\bf b}$.

\item At level $k=1,\ldots,n-1$, (\ref{eq:sumallvector_FTMFT}) becomes
\begin{eqnarray}\label{eq:sumallvector_NONFASTTMFT}
\sum_{{\bf b}} (w_H({\bf v_b})-1)&=&2^{n-k} \sum_{w=0}^k  {k \choose w} (2^{w}-1) \nonumber\\
&=&  2^{n-k}(3^k -2^k)~.
\end{eqnarray}
and the corresponding additions count is given by $\sum_{k=1}^{n-1} 2^{n-k}(3^k -2^k) - (2^k-1)$,
since we have $2^k-1$ separate sums to compute the $2^k-1$ elements of the first row.

Note that (\ref{eq:sumallvector_NONFASTTMFT}) has an extra $2^{n-k}$ scaling factor, when compared to (\ref{eq:sumallvector_FTMFT}).
As observed in the proof of Lemma \ref{fastTMFT}, $\hat{f}_k$ of TMFT is the linear combination of the $2^n$ matrices $E_{\bf b}$, weighted by the scalar values.
Among all the $2^n$ matrices $E_{\bf b}$, there are $2^k$ distinct ones and $2^{n-k}$ repetitions of each distinct one, which causes the extra scaling factor in (\ref{eq:sumallvector_NONFASTTMFT}).

Note that, for TMFT, there is no partial
sum of the time domain samples in (\ref{partI_FT}) and (\ref{partII_FT}),
and thus no extra addition count of $2^k+1$ in (\ref{eq:C2_FTMFT}).
Then, the complexity in (\ref{eq:C2_FTMFT}) becomes
\begin{equation}\label{eq:C2_NONFastTMFT}
\mbox{K}_2 =\sum_{k=1}^{n-1} 2^{n-k}(3^k -2^k) - (2^k-1),~~ k=1,\ldots,n-1~.
\end{equation}

\item At level $0$, as discussed in the fast TMFT, no extra computation complexity is needed, since $f_0$ is already available at level $n$.
Hence, the final complexity of TMFT is
\[
C_{\mbox{TMFT}}=\mbox{K}_1+\mbox{K}_2= 3^{n+1} - (n+4) 2^n + n + 1~.
\] ~$\hfill\square$
\end{enumerate}

\begin{myremark}
Comparing the complexity of TMFT in (\ref{ComplexitynonfastTMFT}) and the fast TMFT in (\ref{ComplexityTMFT2}), we obtain
the asymptotic ratio of $C_{\mbox{TMFT}}$ over $C_{\mbox{FTMFT}}$ as
\begin{equation}
\lim_{n\rightarrow \infty} \frac{C_{\mbox{TMFT}}}{C_{\mbox{FTMFT}}} = 2
\end{equation}
$\hfill\square$
\end{myremark}

\begin{myremark}
We can now compare the complexity of a convolution in the time domain to the complexity when using the fast TMFT.
The convolution in Definition \ref{def:conaddgroup} requires $|G|^2=4^n$ {\em multiplications} in the ring ${\cal R}$. On the other hand, if we apply the convolution theorem, we need to compute two fast TMFT's and one ITMFT for a total of
\[
\frac{3}{2}\left(3^{n+1}-2^{n+2}+1\right) +n2^n
\]
{\em additions} in the ring ${\cal R}$.
$\hfill\square$
\end{myremark}  }

%
\section{Conclusions}\label{secconl}
%
In this paper we have defined the two-modular Fourier transform of a binary function
$f:G\rightarrow {\cal R}$ over $G=C_2^n$ with values in a finite commutative ring
${\cal R}$ of characteristic $2$. This new Fourier transform is based on $k$-dimensional representations
of a sequence of nested subgroups $H_k=C_2^k$ of $G$.
Using the specific group structure of $G$, we have highlighted
the steps that lead to the fast version of the two-modular Fourier transform and its inverse.
In particular, this new inverse Fourier transform significantly deviates from the traditional modular
inverse Fourier transform, which is only valid for the case where the characteristic of the ring
${\cal R}$ does not divide the order of the group $G$.
The major difference is that the trace operator is replaced by a new operator, which
extracts the top right corner element of a matrix.

We then provided the TMFT properties including linearity, shifting property and the convolution theorem, which
enables to efficiently compute convolutions
(multiplications in the group ring ${\cal R}[G]$).
We also presented the exact complexity of fast TMFT and its inverse.


This Fourier transform may have broad applications to problems, where binary functions
need to be reliably computed or in classification of binary functions.
\appendix
\counterwithin{mydefinition}{section}
\setcounter{mydefinition}{0}
\subsection{Basic Definitions of Group Representation and Characters}

\begin{mydefinition}
An $n$--dimensional representation of a group $G$ is a group homomorphism from $G$
to the group of $n \times n$ invertible matrices over a field $K$, i.e.,
\[
 \rho: G \rightarrow GL(n, K)
\]
such that
\[
 \rho(g_1 g_2)=\rho(g_1) \rho(g_2) ~~~~ {\forall} g_1,g_2 \in G~.
\]
If the homomorphism is injective, we say the representation is {\em faithful}.
We also define the {\em kernel} of $\rho$ as $\mbox{Ker}(\rho) = \{g\in G~:~\rho(g) = {\bf I}_n\}$.
$\hfill \square$
\end{mydefinition}
Note that this homomorphism transforms the group operation on a pair of elements to matrix multiplication
of the corresponding representation matrices. Since matrix multiplication is non-commutative these
representations are useful to study non-Abelian groups. When dealing with Abelian groups scalar
(one-dimensional) representations are commonly used \cite{Snaith2003}.
\begin{mydefinition} \label{def:ineqrep}
Given two representations of a group $G$
\[
\rho_1: G \rightarrow GL(n, K)~~~~\rho_1 (g)={\bf V}_g
\]
and
\[
\rho_2: G \rightarrow GL(n, K)~~~~\rho_2 (g)={\bf W}_g
\]
where $g\in G$, we say
$\rho_1$ and $\rho_2$ are {\em equivalent}, if there exists an invertible matrix $\bf{A}$ such that
$\rho_2 (g)={\bf A}\cdot {\bf V}_g \cdot  {\bf A}^{-1} = {\bf W}_g$, for all $g\in G$.
Otherwise, we say $\rho_1$ and $\rho_2$ are {\em inequivalent}.
In the  scalar case, two representations are equivalent only if they coincide, i.e., $\rho_1 (g) =\rho_2 (g)$
for all $g\in G$. $\hfill \square$
\end{mydefinition}
\begin{mydefinition} \label{def:irredrep}
A finite dimensional complex representation $\rho: G \rightarrow GL(n,\mathbb{C})$   is {\em irreducible} if
the only subspace $V \subseteq  \mathbb{C}^n$ that is invariant under all the matrix transformations
$\rho(g)$, for all $g\in G$, is either $V= \mathbb{C}^n$ or $V=0$. $\hfill \square$
\end{mydefinition}
\begin{mydefinition}
Given a representation $\rho$ of a group $G$, the {\em character} of $\rho$ is the function
$\chi_\rho: G \rightarrow K$ given by
\[
\chi_\rho(g) = \mbox{Tr}(\rho(g))  ~~~~ {\forall} g \in G
\]
where Tr$(\cdot)$ is the trace of the matrix.
$\hfill \square$
\end{mydefinition}
{\color{black} Note that a one-dimensional representation coincides with its character and hence it is a group homomorphism.
However, in general the character of a matrix representation is
not a group homomorphism.}

\subsection{Proof of Lemma 3}
According to (\ref{eq:tau_k_u}), for any $g, w, g+w\in G$, we have
\begin{equation}\label{eq:tau_k_u1}
\tau_k (g) = \left\{ \begin{array}{cl}
\sigma_k (u_1) &  ~~{\mbox{if}}~~ g=u_1 + v_1 ~\mbox{for~some}~ \\
&~~u_1 \in H_k/\langle d_k\rangle ~\mbox{and}~{v_1}\in G/{H_k} \\
\overline{\sigma_k (u_1)} & ~~{\mbox{if}}~~ g=u_1 + v_1 + d_k ~\mbox{for~some}~ \\
&~~u_1 \in H_k/\langle d_k\rangle~\mbox{and}~{v_1}\in G/{H_k}
\end{array} \right.
\end{equation}
\begin{equation}\label{eq:tau_k_u2}
\tau_k (w) = \left\{ \begin{array}{cl}
\sigma_k (u_2) &  ~~{\mbox{if}}~~ w=u_2 + v_2 ~\mbox{for~some}~ \\
&~~u_2 \in H_k/\langle d_k\rangle ~\mbox{and}~{v_2}\in G/{H_k} \\
\overline{\sigma_k (u_2)} & ~~{\mbox{if}}~~ w=u_2 + v_2 + d_k ~\mbox{for~some}~ \\
&~~u_2 \in H_k/\langle d_k\rangle~\mbox{and}~{v_2}\in G/{H_k}
\end{array} \right.
\end{equation}
\begin{equation}\label{eq:tau_k_u3}
\tau_k (g+w) \hspace{-1mm}=\hspace{-1mm} \left\{ \begin{array}{cl}
\sigma_k (u_1+u_2) & \hspace{-5mm} \begin{array}{cl}
                        &{\mbox{if}}~g+w=(u_1 + u_2)\hspace{-1mm}+\hspace{-1mm}(v_1+v_2)  \\
                        &\mbox{for~some}~ \\
                        &u_1,u_2, (u_1 + u_2)\in H_k/\langle d_k\rangle~ \\
                        &\mbox{and}~{v_1},{v_2},{v_1+v_2}\in G/{H_k}
                      \end{array}   \\~ \\
\overline{\sigma_k (u_1+u_2)} & \hspace{-5mm}\begin{array}{cl}
&{\mbox{if}}~g+w=(u_1 + u_2) \\ &\hspace{2cm}+(v_1+v_2)+d_k\\
&\mbox{for~some}~ \\
&u_1,u_2, (u_1 + u_2)\in H_k/\langle d_k\rangle~ \\
&\mbox{and}~{v_1},{v_2},{v_1+v_2}\in G/{H_k} \end{array}   \\
\end{array} \right.
\end{equation}
There are four combinations of $g$ and $w$ that we will analyze to prove it is a group homomorphism.
\begin{enumerate}
\item When $g= u_1 + v_1$ and $w=u_2 + v_2$ for some $u_1, u_2 \in H_k/\langle d_k\rangle$ and $~{v_1, v_2}\in G/{H_k}$, we have $g+w=(u_1 + u_2) + (v_1+v_2)$, for some $u_1,u_2, (u_1 + u_2) \in H_k/\langle d_k\rangle$ and ${v_1},{v_2},{v_1+v_2}\in G/{H_k}$.
From (\ref{eq:tau_k_u3}), we have
\begin{equation}\label{eq:case1_tau_k_g+w}
\tau_k (g+w) = \sigma_k (u_1 + u_2)~.
\end{equation}
On the other hand, based on (\ref{eq:tau_k_u1}), (\ref{eq:tau_k_u2}) and Lemma~\ref{lemma:sigma_k}, we have
\begin{eqnarray}\label{eq:case1_tau_k_g+tau_k_w}
\tau_k (g) + \tau_k (w)= \sigma_k (u_1) + \sigma_k (u_2) = \sigma_k (u_1 +u_2)
\end{eqnarray}
Comparing (\ref{eq:case1_tau_k_g+w}) and (\ref{eq:case1_tau_k_g+tau_k_w}),
we have $\tau_k (g+w)= \tau_k (g) + \tau_k (w)$.
\item When $g=u_1 + v_1 +d_k$, and $w=u_2 + v_2 +d_k$ for some
$u_1, u_2 \in H_k/\langle d_k\rangle$ and ${v_1,v_2}\in G/{H_k}$, we have $g+w=(u_1 + u_2) + (v_1+v_2)$
for some $u_1,u_2, (u_1 + u_2) \in H_k/\langle d_k\rangle$ and ${v_1},{v_2},{v_1+v_2}\in G/{H_k}$. From (\ref{eq:tau_k_u3}) and Lemma~\ref{lemma:sigma_k}, we have
    \begin{equation}\label{eq:case2_tau_k_g+w}
    \tau_k (g+w) = \sigma_k (u_1 + u_2) = \sigma_k (u_1) + \sigma_k (u_2)
    \end{equation}
   and
     \begin{eqnarray}\label{eq:case2_tau_k_g+tau_k_w}
    \tau_k (g) + \tau_k (w)&= &\overline{\sigma_k (u_1)} + \overline{\sigma_k (u_2)} \nonumber\\
    &=& {\bf 1}_k + \sigma_k (u_1) + {\bf 1}_k + \sigma_k (u_2)\nonumber\\
    &=& \sigma_k (u_1) +\sigma_k (u_2)
    \end{eqnarray}
    Comparing (\ref{eq:case2_tau_k_g+w})
    and (\ref{eq:case2_tau_k_g+tau_k_w}), we have $\tau_k (g+w)= \tau_k (g) + \tau_k (w)$.
\item When $g=u_1 + v_1$ and $w=u_2 + v_2 +d_k$ for
some $u_1, u_2 \in H_k/\langle d_k\rangle$ and ${v_1,v_2}\in G/{H_k}$, we have $g+w=(u_1 + u_2) + (v_1+v_2) + d_k$,
for some $u_1,u_2, (u_1 + u_2) \in H_k/\langle d_k\rangle$ and ${v_1},{v_2},{v_1+v_2}\in G/{H_k}$. We have
     \begin{eqnarray}\label{eq:case3_tau_k_g+w}
    \tau_k (g+w) &=& \overline{\sigma_k (u_1 + u_2)} \nonumber\\
    &=& {\bf 1}_k + \sigma_k (u_1 + u_2) \nonumber\\
    &=& {\bf 1}_k + \sigma_k (u_1) + \sigma_k (u_2)
    \end{eqnarray}
    and
    \begin{eqnarray}\label{eq:case3_tau_k_g+tau_k_w}
    \tau_k (g) + \tau_k (w) &=& \sigma_k (u_1) + \overline{\sigma_k (u_2)}\nonumber\\
    &=& {\bf 1}_k + \sigma_k (u_1) + \sigma_k (u_2)~.
    \end{eqnarray}
    Comparing (\ref{eq:case3_tau_k_g+w}) and (\ref{eq:case3_tau_k_g+tau_k_w}), we have $\tau_k (g+w)= \tau_k (g) + \tau_k (w)$.
\item When $g=u_1 + v_1 +d_k$ and $w=u_2 + v_2$, for some $u_1, u_2 \in H_k/\langle d_k\rangle$ and ${v_1,v_2}\in G/{H_k}$,
we have $g+w=(u_1 + u_2) + (v_1+v_2) + d_k$, for some $u_1,u_2, (u_1 + u_2) \in H_k/\langle d_k\rangle$ and $~{v_1},{v_2},{v_1+v_2}\in G/{H_k}$.
We obtain the same result as the previous case by swapping $g$ and $w$.
\end{enumerate}
This proves $\tau_k$ to be group homomorphism.
According to the {\color{black} fundamental homomorphism theorem}, we have $\mbox{Ker} (\tau_k) = G/H_k$.
$\hfill\square$

%
\section*{Acknowledgement}
%
We thank Dr Lakshmi Natarajan for fruitful discussions and the anonymous reviewers for their valuable comments.

\bibliographystyle{IEEE}

{\small

\end{document}